\documentclass[11pt,english]{article}
\usepackage{lmodern}

\usepackage[T1]{fontenc}
\usepackage[utf8]{inputenc}
\usepackage{geometry}
\usepackage{amsmath}
\usepackage{graphicx}
\usepackage{setspace}
\makeatletter

\providecommand{\tabularnewline}{\\}

\newcommand{\lyxaddress}[1]{
	\par {\raggedright #1
	\vspace{1.4em}
	\noindent\par}
}

\usepackage[winfonts,UTF8]{ctex}
\usepackage{slashed}
\usepackage{hyperref}
\usepackage{graphicx}
\usepackage[numbers,sort&compress]{natbib}
\date{}

\hypersetup{
colorlinks=false,
linkcolor=red
}

\makeatother

\usepackage{babel}
\begin{document}

\title{Hawking radiation for scalar fields by Einstein-Gauss-Bonnet-de Sitter black holes}

\author{Peng-Cheng Li$^1$\thanks{lipch7@mail.sysu.edu.cn} and Cheng-Yong Zhang$^2$\thanks{ zhangchengyong@fudan.edu.cn}}
\maketitle

\lyxaddress{\begin{center}
\textit{$^1$ School of Physics and Astronomy, Sun Yat-sen University, Zhuhai 519082, China\\\vspace{1mm}
$^2$ Department of Physics and Center for Field Theory and Particle
Physics, Fudan University, Shanghai 200433, China}
\par\end{center}}
\begin{abstract}
We study the greybody factor and power spectra of Hawking radiation
for the minimally or nonminimally coupled scalar field with exact
numerical method in spherically symmetric Einstein-Gauss-Bonnet-de Sitter
black hole spacetime. The effects of scalar coupling constant, angular
momentum number of scalar, spacetime dimension, cosmological constant and
Gauss-Bonnet coupling constant on the Hawking radiation are studied
in detail. Specifically, the Gauss-Bonnet coupling constant always increases the
greybody factor in the entire energy regime. Different from the case of Schwarzschild-de Sitter black
hole, the effects of the scalar
coupling constant on the greybody factor are not monotonic  but relevant to the values of Gauss-Bonnet
coupling constant. Moreover, both these two coupling constants always suppress the power spectra of Hawking radiation in the whole energy regime.
\end{abstract}

\section{Introduction}

The Einstein-Gauss-Bonnet (EGB) gravity is one of the most promising
candidates for modified gravity \cite{Clifton2011}. This theory is a special case of the Lovelock gravity which
is the natural generalization of general relativity in higher dimensions
\cite{Lovelock1971}. It has
second order equations of motion and is ghost free \cite{Zwiebach1985}.
Its action with a positive cosmological constant $\Lambda$
in $d$-dimensional spacetime is
\begin{equation}
S_{G}=\frac{1}{16\pi G}\int d^{d}x\sqrt{-g}\left[R+\alpha\mathcal{L}_{GB}-2\Lambda\right],\label{eq:EGBAction}
\end{equation}
 where $G$ is the $d$-dimensional Newton's constant, $\alpha$ is
the Gauss-Bonnet (GB) coupling constant of dimension $(length)^{2}$,
$R$ is the Ricci scalar, and $\mathcal{L}_{GB}=R_{\mu\nu\rho\sigma}R^{\mu\nu\rho\sigma}-4R_{\mu\nu}R^{\mu\nu}+R^{2}$
is the so-called GB term. The GB term reduces to a topological
surface term in four dimension and is only dynamical
in higher dimensions. It appears as the leading order correction to the low
energy effective action of the heterotic string theory \cite{Boulware1985},
in which $\alpha$ is inversely proportional to the tension of string.

{The GB term is studied not only from the theoretical viewpoint, but also from  phenomenological purpose. As suggested by some extra-dimension
models \cite{Barrau2004,Kanti2004}, the GB coupling constant may be measured by particle colliders through the Hawking radiation of tiny black holes \cite{Dimopoulos2001,Dai2008,Kanti2009} or in high energy cosmic-ray interactions \cite{Feng2001,Emparan2002}.}
In recent years, significant number
of works about the Hawking radiation in higher dimensional spacetime have been done, see the reviews \cite{Kanti2004D,Harmark2007d,Kanti2014Hawking}. It was shown furthermore that Hawking radiation is related to quasinormal modes and superradiance \cite{York1983,Dias2007,Berti:2009kk,Brito:2015oca}. For more recent works, one can see
\cite{Kanti2014,Pappas2016,Kanti2017,RPappasb,bJorge2015, bDong2015,bAhmed2016,bPanotopoulos2017,bKuang2017,Zhang:2014kna,Zhang:2015jda}.

{It becomes more interesting when considering the Hawking radiation in asymptotic de Sitter (dS) spacetime. For minimally or nonminimally coupled massless scalar, the greybody factors in asymptotically flat spacetime tend to zero in
the zero-frequency limit for the waves of arbitrary angular quantum number
$l$ \cite{Page1976,Das1997, Higuchi2002,Harris2003,Chen2010}.
However, the presence of a positive cosmological constant leads to different results.
The greybody factor for the $l=0$  mode of a minimally coupled massless scalar by the Schwarzschild-de Sitter (SdS) black holes
does not vanish in the infrared limit \cite{Kanti2005,Brady1997}.}
This is due to that the zero-frequency particles are fully
delocalized, and have therefore a finite probability to traverse the distance between the two horizons \cite{Kanti2005}. When the scalar is massive or equivalently the coupling is nonminimal,   the infrared enhancement
of transmitted flux in SdS will not present \cite{Crispino2013}.

For the case of EGB gravity, the works on Hawking radiation are much less.
For scalar and graviton emissions, the numerical studies of the GB
black hole in an asymptotic flat spacetime were carried out in \cite{Grain2005, Konoplya2010}. Here we would like to mention that a new  approximate method was adopted to obtain the analytic  expressions for the greybody factors
of nonminimally scalar fields by a higher dimensional
Schwarzschild-dS (SdS) black hole \cite{Kanti2014}. By using the same method  the study  was extended to the case of EGB-dS black holes in \cite{Zhang2017}. The properties of the greybody factors and power spectra of Hawking radiation in terms of the particle and spacetime
parameters were analyzed  in detail. It was
found that the nonminimal coupling suppresses the greybody factor and
the GB coupling enhances the greybody factor, but  they both
suppress the energy emission rate of Hawking radiation. However, it
should be pointed out that these results were obtained  under the assumption that both the cosmological constant and the
nonminimal coupling constant are small. Moreover, the results become problematic in the high frequency regime or in odd dimensional spacetime. In this paper, we will remedy
these defects by
employing exact numerical method to study the Hawking radiation of EGB-dS black holes in the entire energy regime without any approximation imposed on the parameters. The previously obtained analytical solutions
will be served as asymptotic boundary conditions here. We will compare
the numerical results with the approximate analytical ones and analysis the deviations stemming from various parameters. Our study is similar to the case of SdS black holes in \cite{Pappas2016}.

The outline of our paper is as follows. In Sec. \ref{sec:Background}, we present the basics of EGB-dS
black hole and the equation of motion for the scalar field. In Sec. \ref{sec:The-numerical-solution},
we describe the numerical method solving the perturbation equation.
Section \ref{sec:Greybody-factor} gives the numerical results for the
greybody factor and Sec. \ref{sec:Energy-emission-rate} gives the
numerical results for energy emission rate of Hawking radiation, on which the
effects of various parameters are analyzed in detail. We summarize our results
in Sec. \ref{sec:Summary}.

\section{Background\label{sec:Background}}

The $d$-dimensional spherically symmetric EGB-dS black hole solution
of the theory (\ref{eq:EGBAction}) is described by \cite{Zwiebach1985}
\begin{align}
ds^{2} & =-hdt^{2}+\frac{dr^{2}}{h}+r^{2}d\Omega_{d-2}^{2},\label{eq:4metric}\\
h & =1+\frac{r^{2}}{2\tilde{\alpha}}\left(1-\sqrt{1+\frac{4\tilde{\alpha}m}{r^{d-1}}+\frac{8\tilde{\alpha}\Lambda}{(d-1)(d-2)}}\,\right).\nonumber
\end{align}
 Here $d\Omega_{d-2}^{2}$ is the line element of the $(d-2)$-dimensional
unit sphere $S^{d-2}$. $\tilde{\alpha}$ is related to the GB
coupling constant by $\tilde{\alpha}=\alpha(d-3)(d-4)$. The parameter
$m$ is related to the black hole mass $M$ by $m=\frac{16\pi GM}{(d-2)\Omega_{d-2}}$. Depending on the parameters $M$, $\Lambda$ and $\tilde{\alpha}$, the black solution may have two, one or zero horizons. In this paper, we will choose parameters such that the
EGB-dS black hole spacetime always has two horizons: the event horizon
$r_{h}$ and cosmological horizon $r_{c}$. Moreover, it was found that the EGB-dS black holes
could be  unstable in certain parameter regime. In our discussions,
the parameters are restricted to the stable regime
\cite{Konoplya2008,Cuyubamba2016,Konoplya2017}.
In terms of the black hole horizon radius $r_{h}$, the mass parameter $m$ can
be expressed as
\begin{equation}
m=r_{h}^{d-3}\left(1+\frac{\tilde{\alpha}}{r_{h}^{2}}-\frac{2\Lambda r_{h}^{2}}{(d-1)(d-2)}\right),
\end{equation}
so that $m$ can be eliminated from the following discussions.

We consider a massless scalar field $\Phi$ propagating in  the  aforementioned  EGB-dS spacetime. The scalar field is coupled minimally or nonminimally to gravity, and the corresponding action of the scalar part is
\begin{equation}
S_{\Phi}=-\frac{1}{2}\int d^{d}x\sqrt{-g}[\xi\Phi^{2}R+\partial_{\mu}\Phi\partial^{\mu}\Phi],
\end{equation}
where $\xi$ is the  nonminimal coupling constant, with $\xi=0$ corresponding to the minimally coupled case. The equation of motion
of the scalar field is
\begin{equation}
\nabla_{\mu}\nabla^{\mu}\Phi=\xi R\Phi. \label{eq:EOMs}
\end{equation}
 Assuming that the effect of $\Phi$ on the background spacetime is
negligible, then the above equation will be solved in a fixed background
given by (\ref{eq:4metric}). For spherically symmetric background,
we can decompose the scalar wave function as
\begin{equation}
\Phi=e^{-i\omega t}\phi(r)Y_{(d-2)}^{l}(\Omega),
\end{equation}
 where $Y_{(d-2)}^{l}(\Omega)$ are spherical harmonics  of the scalar wave function on $S^{d-2}$ with angular
momentum number $l$. The
angular and the radial parts are decoupled such that the radial equation
reads
\begin{equation}
\frac{1}{r^{d-2}}\frac{d}{dr}\left(hr^{d-2}\frac{d\phi}{dr}\right)+\left[\frac{\omega^{2}}{h}-\frac{l(l+d-3)}{r^{2}}-\xi R\right]\phi=0.\label{eq:RadialEq}
\end{equation}
Here the Ricci scalar for metric (\ref{eq:4metric}) is
\begin{equation}
R=-\frac{d^{2}h}{dr^{2}}+\frac{d-2}{r^{2}}\left(-2r\frac{dh}{dr}+(d-3)(1-h)\right).
\end{equation}
 Introducing a new variable $u(r)=r^{\frac{d-2}{2}}\phi(r)$, we get
a Schrödinger-like equation
\begin{equation}
\frac{d^{2}u}{dr_{\star}^{2}}+(\omega^{2}-V(r_{\star}))u=0
\end{equation}
 where $r_{\star}$ is the tortoise coordinate defined by $dr_{\star}=dr/h$.
The effective potential felt by the scalar field is
\begin{equation}
V(r_{\star})=h\left[\frac{l(l+d-3)}{r^{2}}+\xi R+\frac{d-2}{2r}h'+\frac{(d-2)(d-4)}{4r^{2}}h\right].
\end{equation}
The potential vanishes at both the event horizon and cosmological horizon of EGB-dS
black hole, which allows us to analytically derive the greybody factor by using matched asymptotic expansion method \cite{Zhang2017}. In the intermediate region of these two horizons, the potential has the form of a barrier. In \cite{Zhang2017} it was revealed that
 due to the presence of the GB coupling constant $\tilde{\alpha}$,
 the dependence of the profile of the barrier  on both spacetime and particle parameters becomes more subtle. For example, without $\tilde{\alpha}$  the height of the barrier increase with the coupling parameter $\xi$ of the scalar field, however, when $\tilde{\alpha}$ has a large nonzero value,  $\xi$  suppresses the barrier. In the following we will study the effects of the various parameters on the Hawking radiation in detail.

\section{The boundary conditions for numerical integrations \label{sec:The-numerical-solution}}

To get the information of the Hawking radiation, such as the greybody
factor and power spectra of the scalar field, one should solve the radial
equation (\ref{eq:RadialEq}) at first. In \cite{Zhang2017},
by solving  (\ref{eq:RadialEq}) near the event
horizon and cosmological horizon separately and matching them in the
intermediate region, they derived an approximatively analytical formula
for the greybody factors when the cosmological constant and nonminimal coupling constant of the filed are small.
Though it is valid for all partial modes and may hold beyond the low
energy regime, the deviation in the high energy regime is obvious.
Besides, it works only in even dimensional spacetime. In this paper,
we will solve the radial equation numerically to give the exact results
and overcome the drawbacks of the approximatively analytical method.

We now briefly review the main results of the approximatively analytic approach \cite{Zhang2017}.
The radial equation (\ref{eq:RadialEq}) was first solved near the event horizon. By using the coordinate transformation
\begin{equation}
f=\frac{h}{1-\tilde{\Lambda}r^{2}},\label{eq:EventTrans}
\end{equation}
 in which $\tilde{\Lambda}=-\frac{1}{2\tilde{\alpha}}\left(1-\sqrt{1+\frac{8\tilde{\alpha}\Lambda}{(d-1)(d-2)}}\right)$
for EGB-dS black hole and imposing the ingoing boundary condition near the event
horizon, the asymptotic solution has the form
\begin{equation}
\left.\phi(r)\right|_{r\simeq r_{h}}=A_{1}f^{\alpha_{1}}(1-f)^{\beta_{1}}F(a_{1},b_{1},c_{1},f),\label{eq:EventSol}
\end{equation}
where $A_{1}$ is an arbitrary integration constant and $F$ stands for the hypergeometric function.
The expressions for parameters $\alpha_{1},\beta_{1}$ and hypergeometric
indices $(a_{1},b_{1},c_{1})$ can be found in (3.9) and (3.10) in
\cite{Zhang2017}.

On the other hand, near the cosmological horizon  by using the variable
$\tilde{h}\equiv1-\tilde{\Lambda}r^{2}$, the radial equation (\ref{eq:RadialEq}) was solved by
\begin{equation}
\left.\phi(r)\right|_{r\simeq r_{c}}=B_{1}\tilde{h}^{\alpha_{2}}(1-\tilde{h})^{\beta_{2}}F(a_{2},b_{2},c_{2},\tilde{h})+B_{2}\tilde{h}^{-\alpha_{2}}
(1-\tilde{h})^{\beta_{2}}F(1+a_{2}-c_{2},1+b_{2}-c_{2},2-c_{2},\tilde{h}).\label{eq:CosmSol}
\end{equation}
 Here $B_{1},B_{2}$ are integration constant. The expressions for
parameters $\alpha_{2},\beta_{2}$ and hypergeometric indices $(a_{2},b_{2},c_{2})$
are given in (3.16) and (3.18) in \cite{Zhang2017}. Note that
$\tilde{h}$ is an approximation of the metric
function $h=1-\tilde{\Lambda}r^{2}-(r_{h}/r)^{d-3}(1-\tilde{\Lambda}r^{2})$
near the cosmological horizon $r_{c}$, which requires the smallness of
the cosmological constant. The greybody factor for  the emission   of the scalar field by the black hole is determined by the amplitudes of the incoming and outgoing wave at the cosmological horizon. In term of the integration constants it is  given by
\begin{equation}
|\gamma_{\omega l}|^{2}=1-\left|\frac{B_{2}}{B_{1}}\right|^{2}.\label{eq:GreyFactor}
\end{equation}
The ratio $B_2/B_1$ can be analytically obtained from the matching of the two asymptotic solutions (\ref{eq:EventSol}) and (\ref{eq:CosmSol}) at   intermediate regime, which occurs only if both the cosmological constant and the nonminimal coupling constant remain small. One interesting  behavior of the result is that for the  minimal coupling  case and for the $l=0$ mode, the greybody factor is not vanishing even in the zero-frequency limit, similar to the SdS case.
{For instance, when $l=0$, for small $\tilde{\alpha}$ and in the limit $\omega \to 0$,
}
one finds \cite{Zhang2017}
\begin{equation}
|\gamma_{\omega0}|^{2}=\frac{4(r_{h}r_{c})^{d-2}}{\left(r_{h}^{d-2}+r_{c}^{d-2}\right)^{2}}+\frac{4r_{c}^{d-2}r{}_{h}^{d-2}(r_{c}^{d-2}-r_{h}^{d-2})}{(r_{c}^{d-2}+r_{h}^{d-2})^{3}}\frac{\tilde{\alpha}}{r_{h}^{2}}+O(\omega,\tilde{\alpha}^2).\label{eq:ALowLimit}
\end{equation}
Since no extra  approximations are made regarding the parameters, except for the smallness of the cosmological constant and the coupling constant, the above approximate method seems to be powerful, as it covers not only the most parameter regime but also the entire energy regime.
However, as we will see later, its deviations from the exact solution becomes plain when moving to the high energy regime. This entails the use of the exact numerical method to solve the radial equation  (\ref{eq:RadialEq}) without making any approximation for the parameters.

First of all, we  determine the boundary conditions.
Near the event horizon $r\to r_{h}$, we have $f\to0$
and the asymptotic solution \ref{eq:EventSol} becomes
\begin{equation}
\left.\phi(r)\right|_{r\to r_{h}}\simeq A_{1}f^{\alpha_{1}}=A_{1}e^{-i\frac{\omega r_{h}}{A_{h}}\ln f}.\label{eq:EventSolB}
\end{equation}
Here we have used $\alpha_{1}=-i\frac{\omega r_{h}}{A_{h}}$ and $A_{h}=A(r_{h})$
in which
\begin{equation}
A(r)=-2+\frac{d-1}{2}\left[1+\left(1+\frac{4\tilde{\alpha}m}{(1+2\tilde{\alpha}\tilde{\Lambda})^{2}}\frac{1}{r^{d-1}}\right)^{-1/2}\right](1-\tilde{\Lambda}r^{2}).
\end{equation}
 The solution (\ref{eq:EventSolB}) describes an ingoing wave near
the event horizon. The coefficient $A_{1}$ has no specific physical significance so it can be
fixed  by requiring
\begin{equation}
\phi(r_{h})=1.\label{eq:0D}
\end{equation}
 To solve (\ref{eq:RadialEq}) numerically, we still need the first derivative of $\phi$
near the event horizon which is given by
\begin{equation}
\left.\frac{d\phi}{dr}\right|_{r=r_{h}}=\left.A_{1}e^{-i\frac{\omega r_{h}}{A_{h}}\ln f}\left(-i\frac{\omega r_{h}}{A_{h}}\right)\frac{A(r)(1-f)}{hr}\right|_{r=r_{h}}\simeq-i\frac{\omega}{h(r_{h})}.\label{eq:1D}
\end{equation}

Near the cosmological horizon $r\to r_{c}$, the approximation $h\simeq1-\tilde{\Lambda}r^{2}$ will be less accurate  when $r_{c}$ is small or
the cosmological constant is large. To avoid this, in the numerical calculations we still
take $f$ in (\ref{eq:EventTrans}) as the variable.
Following a similar procedure, the asymptotic solution near $r_{c}$
can be worked out as
\begin{equation}
\left.\phi(r)\right|_{r\to r_{c}}=B_{1}f^{\alpha_{2}}+B_{2}f^{-\alpha_{2}}=B_{1}e^{-i\frac{\omega r_{c}}{A_{c}}\ln f}+B_{2}e^{i\frac{\omega r_{c}}{A_{c}}\ln f}\label{eq:SolRc}
\end{equation}
with $\alpha_{2}=-i\frac{\omega r_{c}}{A_{c}}$ in which $A_{c}=A(r_{c})$.
The first and second parts correspond ingoing and outgoing waves,
respectively. Therefore, the
greybody factor is given again by the expression (\ref{eq:GreyFactor})

Provided the  boundary conditions (\ref{eq:0D}) and (\ref{eq:1D}), the radial
equation (\ref{eq:RadialEq}) can be solved numerically from the event horizon.
The numerical integrations proceed to the cosmological horizon, from which the coefficients $B_{1},B_{2}$ can be extracted from (\ref{eq:SolRc})
by using
\begin{align}
B_{1}= & \left.\frac{1}{2}e^{i\frac{\omega r_{c}}{A_{c}}\ln f}\left[\phi+\frac{iA_{c}hr}{\omega r_{c}A(r)(1-f)}\frac{d\phi}{dr}\right]\right|_{r=r_{c}},\label{eq:B12}\\
B_{2}= & \left.\frac{1}{2}e^{-i\frac{\omega r_{c}}{A_{c}}\ln f}\left[\phi-\frac{iA_{c}hr}{\omega r_{c}A(r)(1-f)}\frac{d\phi}{dr}\right]\right|_{r=r_{c}}.\nonumber
\end{align}

\begin{figure}[h]
\begin{centering}
{\footnotesize{}}%
\begin{tabular}{cc}
{\footnotesize{}\includegraphics[scale=0.4]{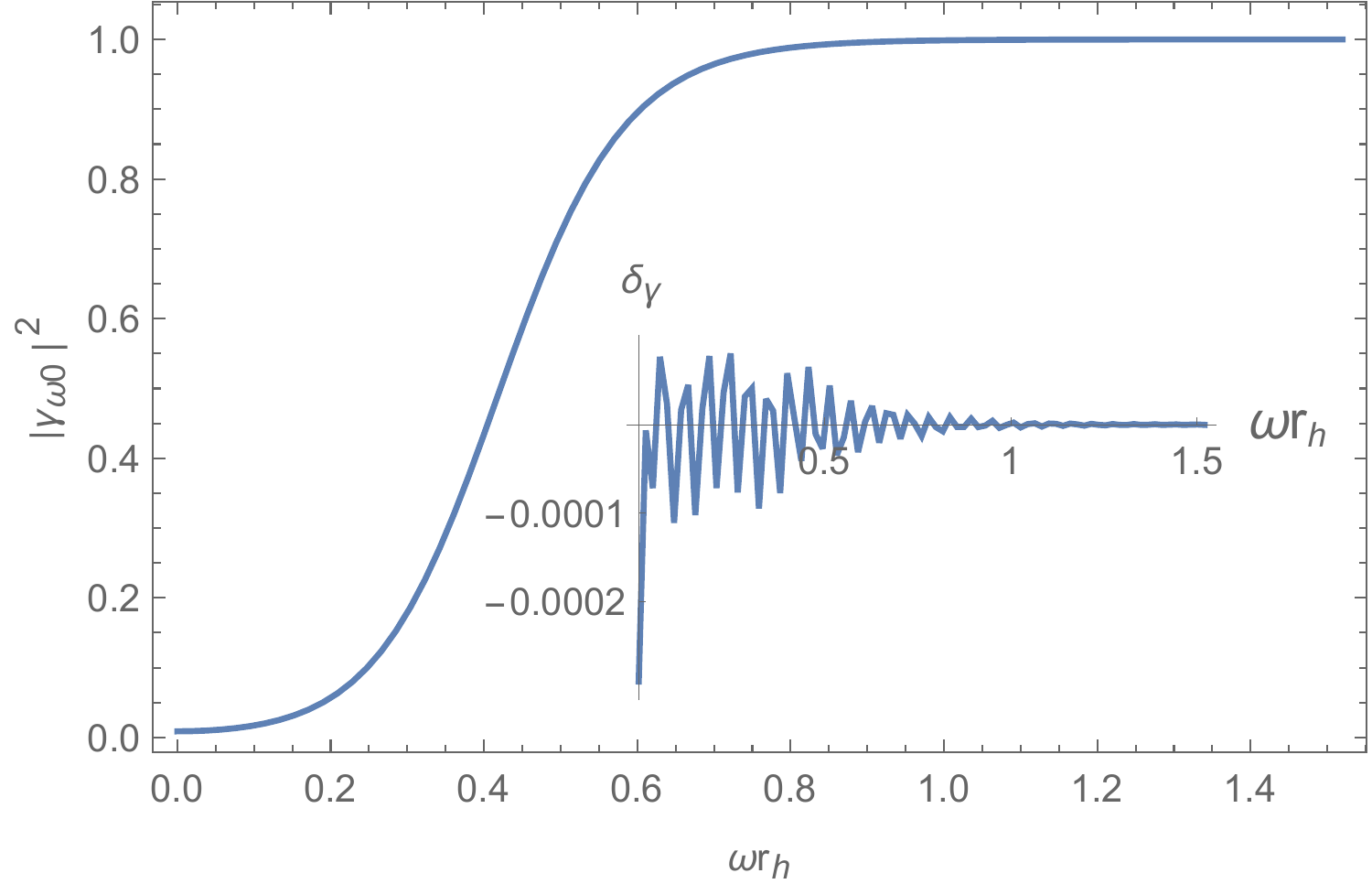}} & {\footnotesize{}\includegraphics[scale=0.4]{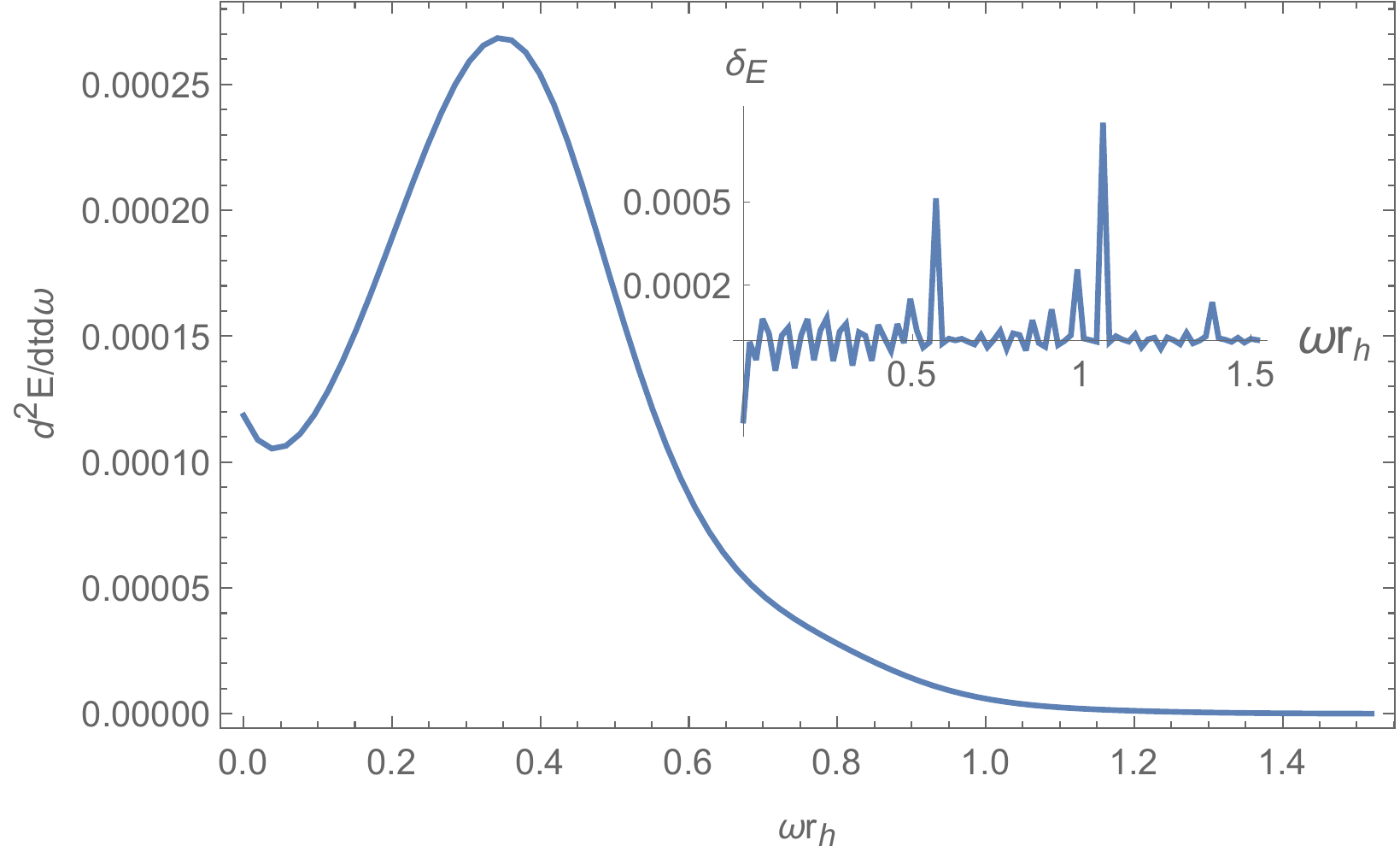}}\tabularnewline
\end{tabular}{\footnotesize\par}
\par\end{centering}
{\footnotesize{}\caption{\label{fig:error} The greybody factor for $l=0$ (left) and power spectra (right) with $d=5$, $\Lambda=0.1$, $\xi=0$, $\tilde{\alpha}=0.6$. The inserts show the
 relative errors of the greybody factor $\delta_{\gamma}=\frac{|\gamma_{\omega0}|_{\epsilon=10^{-5}}^{2}-|\gamma_{\omega0}|_{\epsilon=10^{-8}}^{2}}{|\gamma_{\omega0}|_{\epsilon=10^{-8}}^{2}}$
 and the power spectra $\delta_{E}=\frac{\frac{d^{2}E}{dtd\omega}|_{\epsilon=10^{-5}}-\frac{d^{2}E}{dtd\omega}|_{\epsilon=10^{-8}}}{\frac{d^{2}E}{dtd\omega}|_{\epsilon=10^{-8}}}$
, respectively.}
}{\footnotesize\par}
\end{figure}

{Note that as $r\to r_{h}$, $h(r)$ tends to 0, so that (\ref{eq:1D}) tends
to infinity. $f(r)$ also tends to 0 when $r\to r_{c}$
such that (\ref{eq:B12}) is ill defined.  We should shift the boundary from $r=r_{h}$ to $r=r_{h}+\epsilon$ and  $r=r_{c}$
to $r=r_{c}-\epsilon$
to make the boundary condition manipulable.  The small shift $\epsilon$
will be chosen appropriately to assure the stability of the numerical
results. In Fig. \ref{fig:error} we show the relative error of greybody factor and power spectra between the results when $\epsilon=10^{-5}$ and $\epsilon=10^{-8}$. They are so small   that our numerical results can be trusted. }

{\footnotesize{}}

\section{Greybody factor \label{sec:Greybody-factor}}

\subsection{Comparison of numerical and analytical results}
\begin{figure}[h]
\begin{centering}
{\footnotesize{}}%
\begin{tabular}{cc}
{\footnotesize{}\includegraphics[scale=0.35]{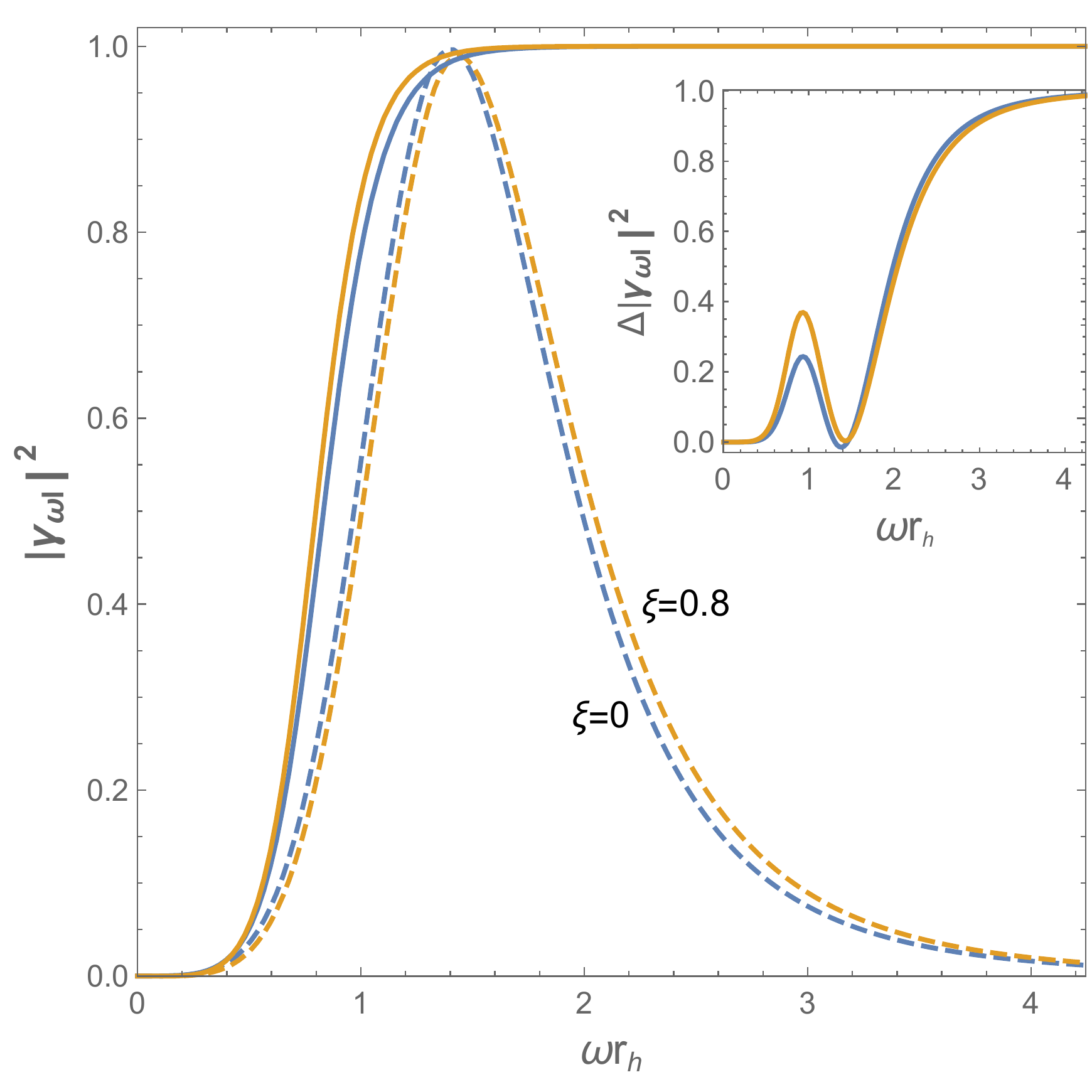}} & {\footnotesize{}\includegraphics[scale=0.35]{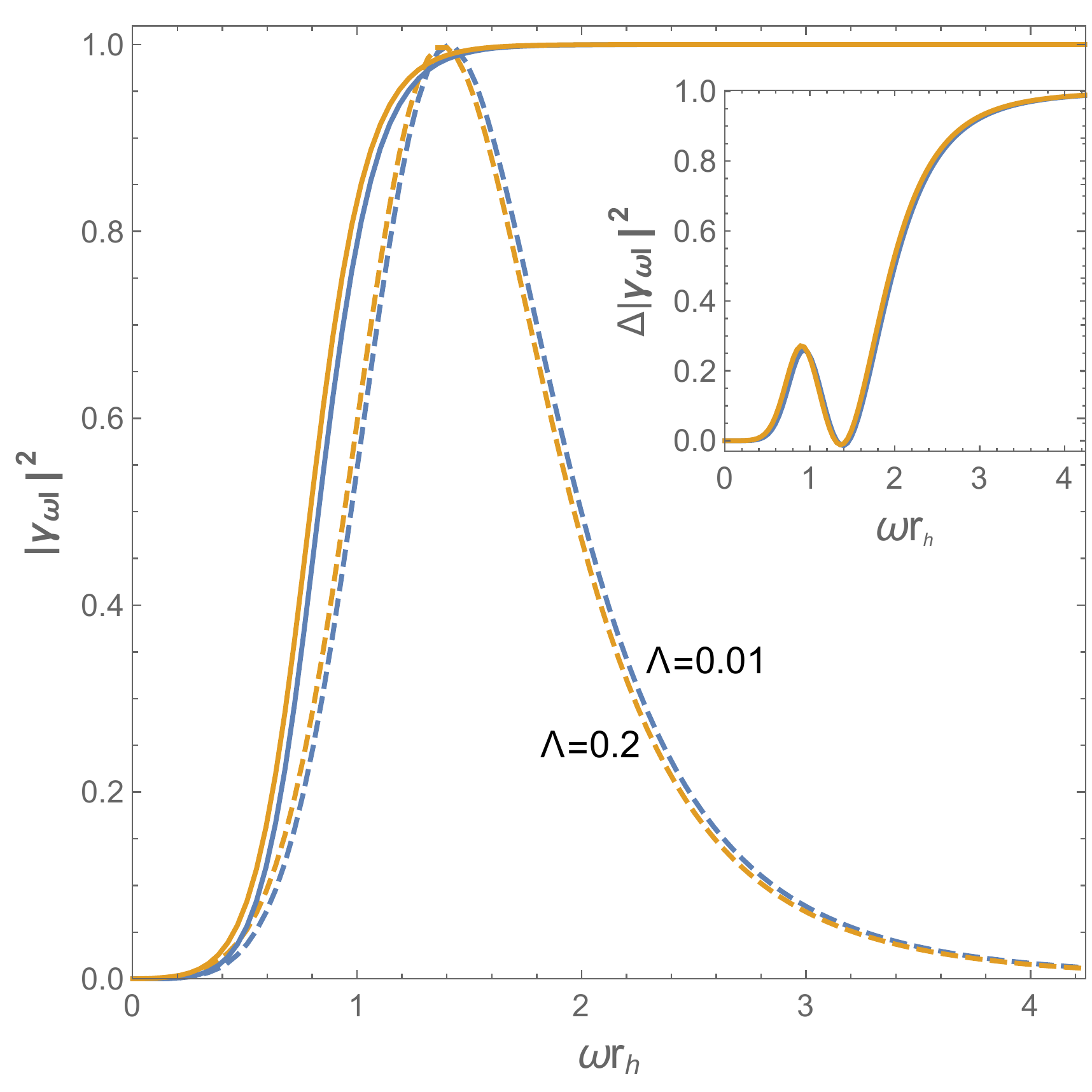}}\tabularnewline
\end{tabular}{\footnotesize\par}
\par\end{centering}
{\footnotesize{}\caption{\label{fig:GreyLxi} Analytical (dashed lines) and numerical (solid
lines) results for the greybody factors when $d=6,l=0,\tilde{\alpha}=0.5$.
Left panel for $\Lambda=0.01$. Blue lines for $\xi=0$, yellow lines
for $\xi=0.8$. Right panel for $\xi=0.1$. Blue lines for $\Lambda=0.01$,
yellow lines for $\Lambda=0.2$. The inserts show the difference of
greybody factors $\Delta|\gamma_{\omega l}|^{2}=|\gamma_{\omega l}|_{numerical}^{2}-|\gamma_{\omega l}|_{analytical}^{2}$
between numerical result and analytical result with the same parameters.}
}{\footnotesize\par}
\end{figure}

With the  numerical result at our disposal, we can compare it with the analytical expression for the greybody factor derived in \cite{Zhang2017}.
In Fig. \ref{fig:GreyLxi}, we plot the numerical and analytical results  for the greybody factors for different values of the  cosmological constant $\Lambda$ and the nonminimal coupling constant $\xi$.
They are in well agreement in the low energy regime. However, the
deviations of the approximatively analytical results from the exact numerical
results become evident as $\xi$ or $\Lambda$ increases. This is
expected since it was assumed small $\xi$ and $\Lambda$ to get
the analytical expression of greybody factors in \cite{Zhang2017}.

On the other hand, the greybody factor should tend to 1 when the energy
is large enough no matter what value $\xi$ or $\Lambda$ takes, as
shown by the numerical result in Fig. \ref{fig:GreyLxi}. This is
due to the fact that the particle with high energy can overcome the
effective potential barrier easily and escape to far region. However, the
analytical result tends to $0$ in the high energy regime and thus become
unreliable at all. We will show only the effects of various parameters on the numerical result hereafter, since it is valid
over the entire energy regime and for
arbitrary values of the particle and spacetime parameters.

{Moreover, the greybody factors for odd dimensional spacetime cannot be obtained from the approximatively analytical method in \cite{Zhang2017}. The numerical method here can remedy this flaw and the effects of dimension will be analyzed in subsection \ref{subsec:dgreybody}. Hereafter, we focus mainly on the odd dimensional spacetime to emphasize the new results.}

 Without the GB term in the theory, the dependence of the greybody factors on the various parameters has been  studied numerically  in \cite{Pappas2016}. The results can be summarized as follows. The greybody factor is
suppressed  by the angular momentum number  $l$, both for minimally or nonminimally coupled scalar. The similar case occurs for the spacetime dimensions $d$.
Moreover, the nonminimal coupling constant $\xi$ decreases the greybody
factor when other parameters are fixed. The effect of the cosmological constant on the greybody factor depends on the value of $\xi$. The cosmological constant enhances the greybody
factor for small $\xi$,  but for large $\xi$ it leads to a suppression in the greybody factor especially
in the low-energy regime. As we will see in the following, the presence of the GB term in the theory makes the
analysis more subtle than the case of general relativity.

{\footnotesize{}}

\subsection{\label{subsec:AlphaL}Effects of $\tilde{\alpha}$ on the greybody
factor for different partial modes $l$}

{\footnotesize{}}
\begin{figure}[h]
\begin{centering}
{\footnotesize{}}%
\begin{tabular}{cc}
{\footnotesize{}\includegraphics[scale=0.3]{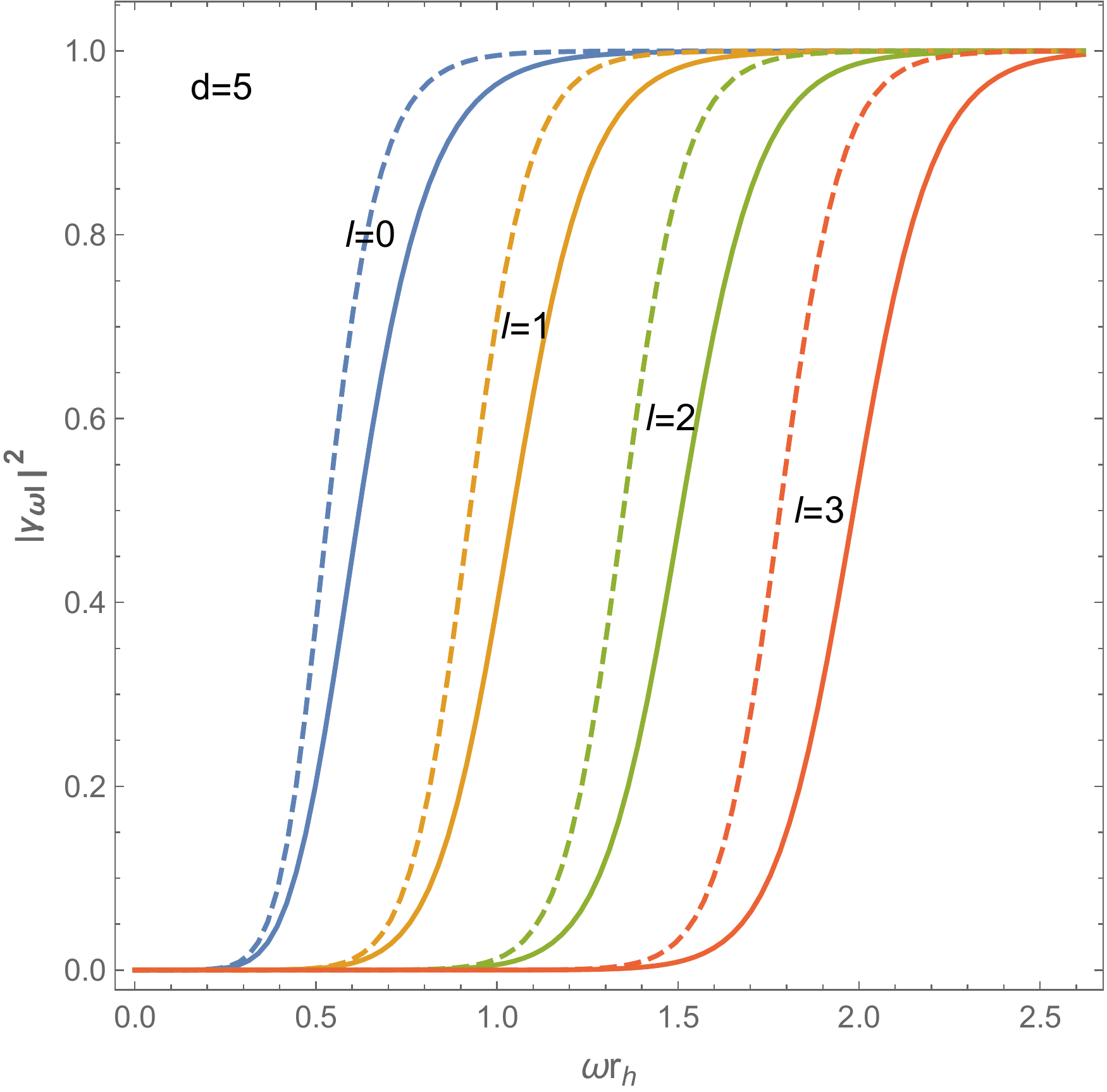}} & {\footnotesize{}\includegraphics[scale=0.3]{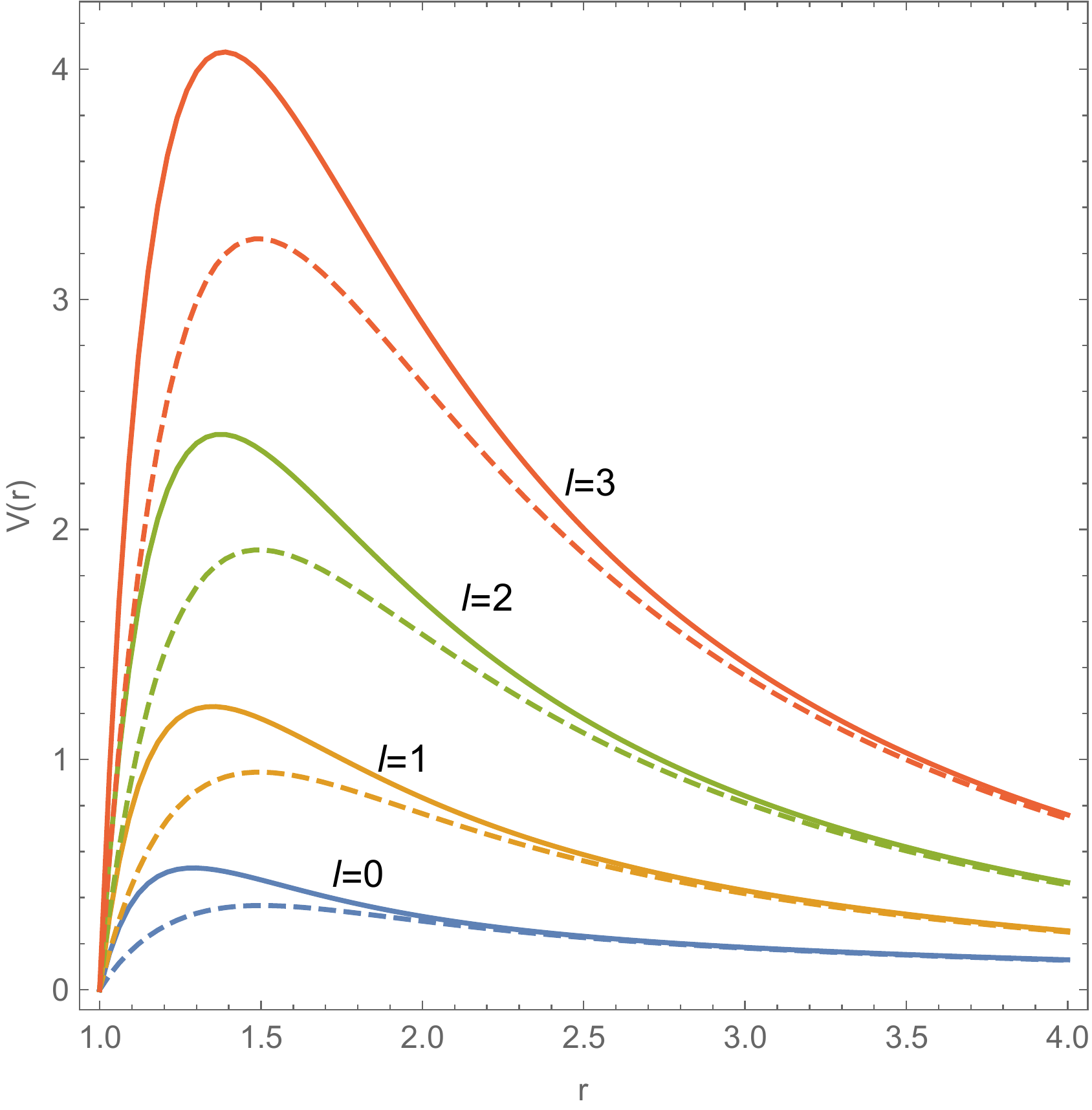}}\tabularnewline
\end{tabular}{\footnotesize\par}
\par\end{centering}
{\footnotesize{}\caption{\label{fig:AlphaL} Effect of $\tilde{\alpha}$ on the greybody factor
for different mode $l$ and the corresponding effective potential.
Solid lines for $\tilde{\alpha}=0$, dashed lines for $\tilde{\alpha}=0.3$.
We fix $d=5$, $\Lambda=0.1$, $\xi=0.6$ here.}
}{\footnotesize\par}
\end{figure}

In this subsection, we study the effect of $\tilde{\alpha}$ on the
greybody factor for different partial modes $l$.
{We take $d=5$ for example. The behaviors in other
dimensions are qualitatively similar.}
From the left panel
of Fig. \ref{fig:AlphaL}, we see that the greybody factor is suppressed
significantly as $l$ increases with other parameters fixed. As a consequence, the
mode $l=0$ dominates the radiation. On the other hand, $\tilde{\alpha}$
increases the greybody factors for all modes $l$ in the whole energy
regime. We find that this qualitative behavior is independent of the
spacetime dimension $d$, scalar coupling $\xi$ and cosmological
constant $\Lambda$. In the right panel, we plot the corresponding
effective potential to understand this behavior intuitively. For  larger $l$, when other parameters are fixed, the wave must transverse
a higher effective potential barrier, thus its transmission is suppressed
with $l$. At the meantime, $\tilde{\alpha}$ always decreases the
effective potential barrier so that it becomes easier for the scalar
to transverse the barrier and the greybody factor increases with $\tilde{\alpha}$.

\subsection{\label{subsec:XiL}Competition between  $\xi$ and $\tilde{\alpha}$}

{\footnotesize{}}
\begin{figure}[h]
\begin{centering}
{\footnotesize{}}%
\begin{tabular}{cc}
{\footnotesize{}\includegraphics[scale=0.3]{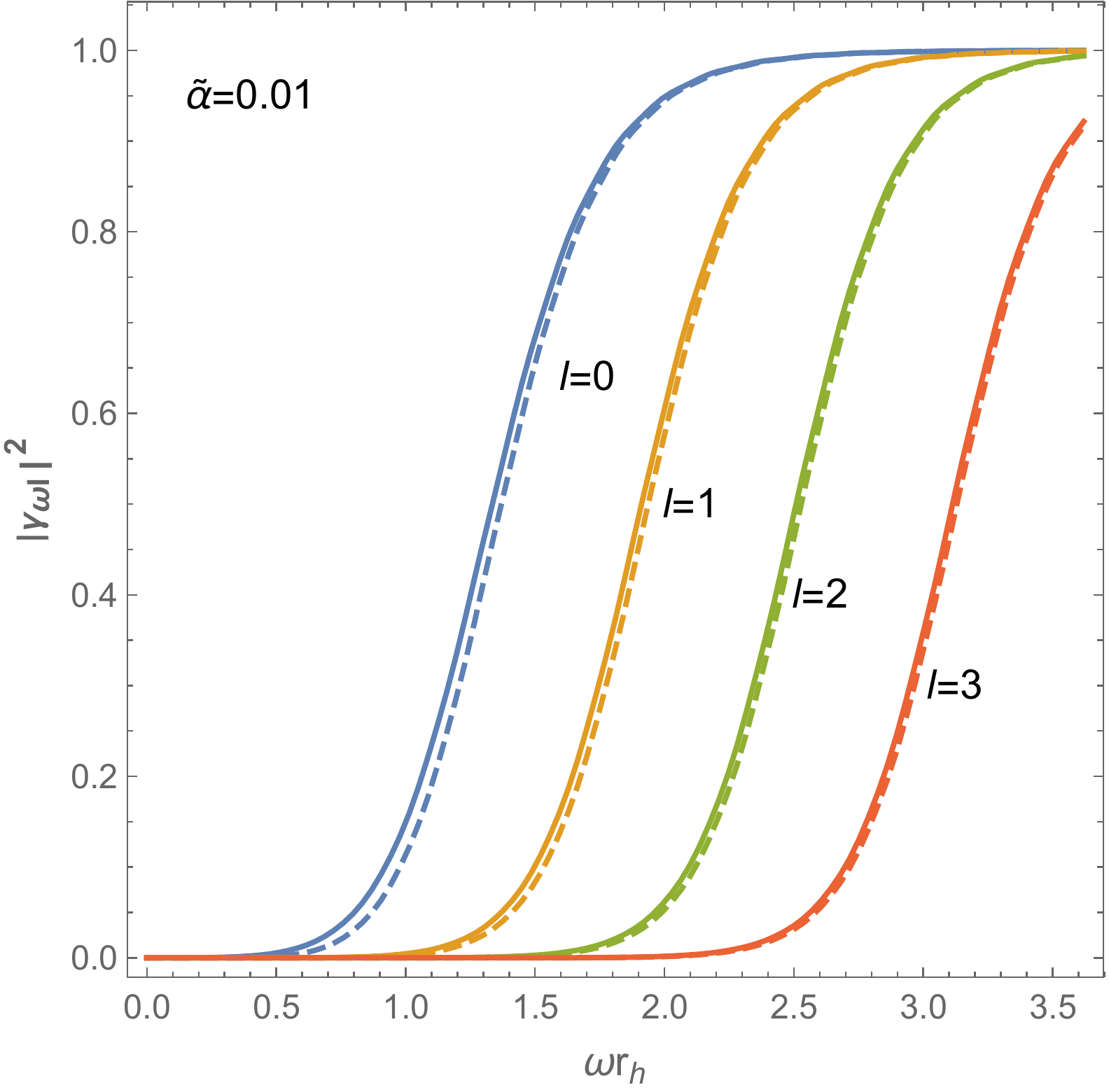}} & {\footnotesize{}\includegraphics[scale=0.3]{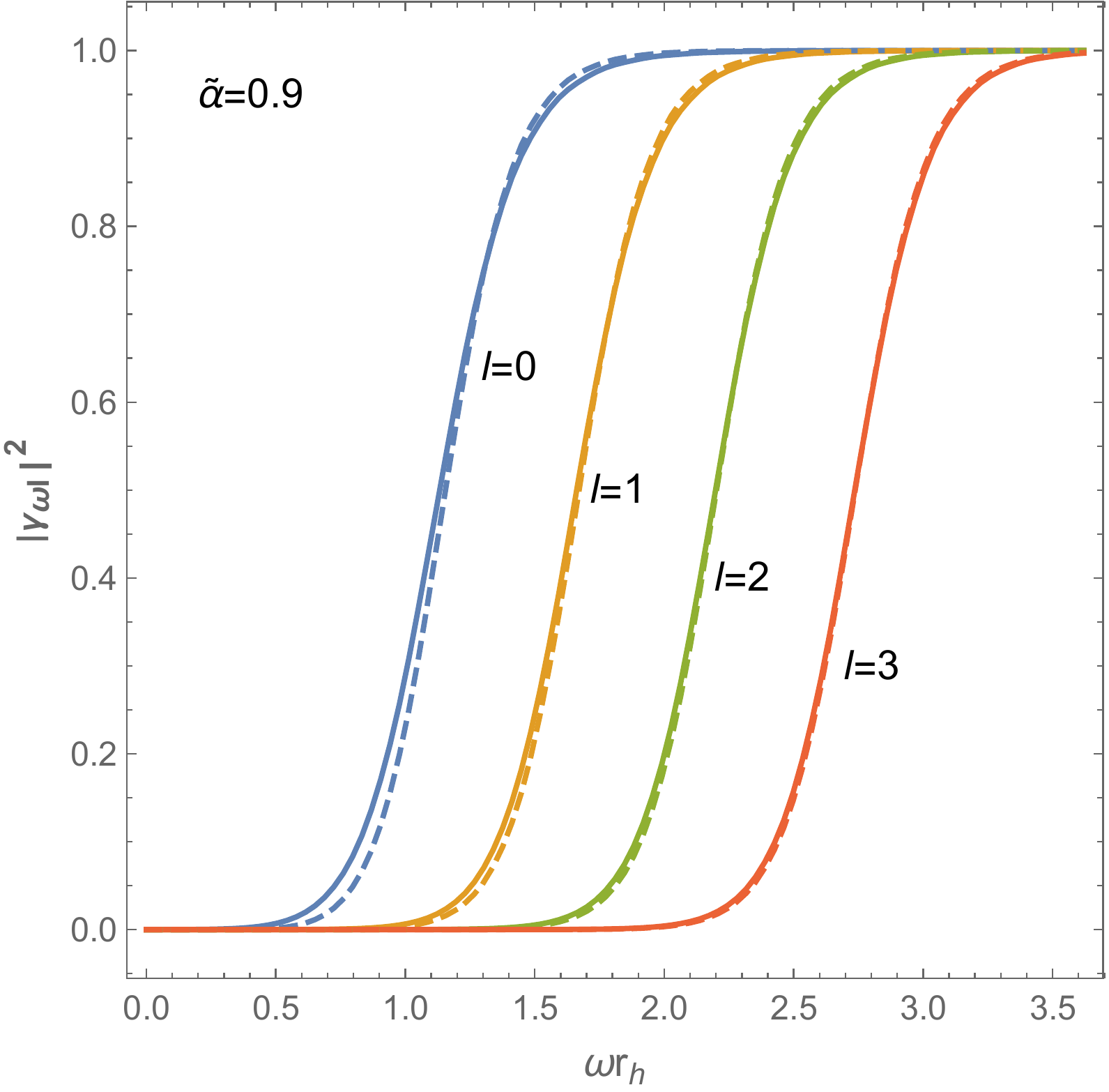}}\tabularnewline
\end{tabular}{\footnotesize\par}
\par\end{centering}
{\footnotesize{}\caption{\label{fig:XiL} Effect of $\xi$ on the greybody factor for different
$l$. Solid lines for $\xi=0,$ dashed lines for $\xi=0.6$. We fix
$d=7,\Lambda=0.1$ here.}
}{\footnotesize\par}
\end{figure}

\begin{figure}[h]
\begin{centering}
{\footnotesize{}}%
\begin{tabular}{cc}
{\footnotesize{}\includegraphics[scale=0.3]{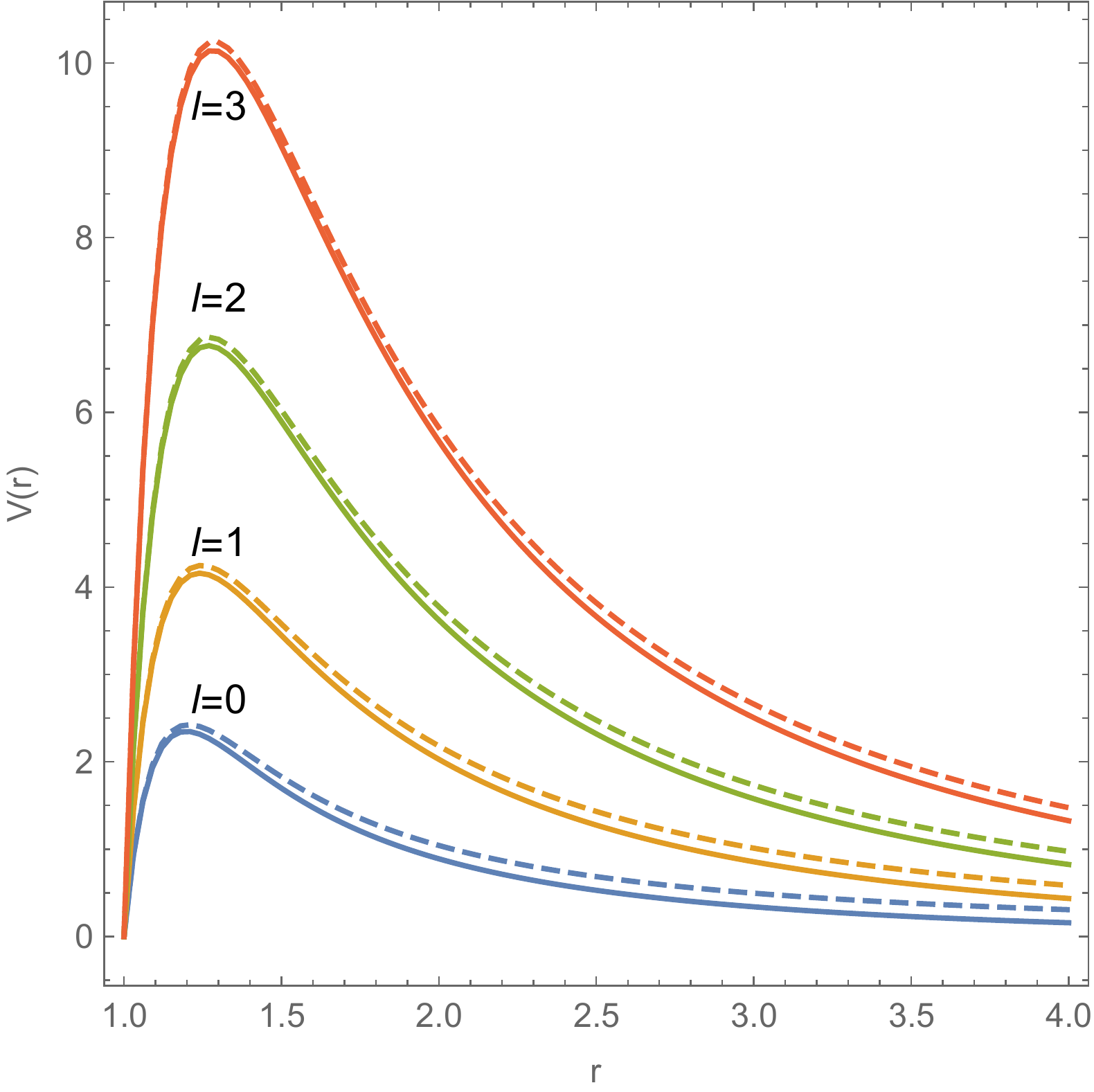}} & {\footnotesize{}\includegraphics[scale=0.3]{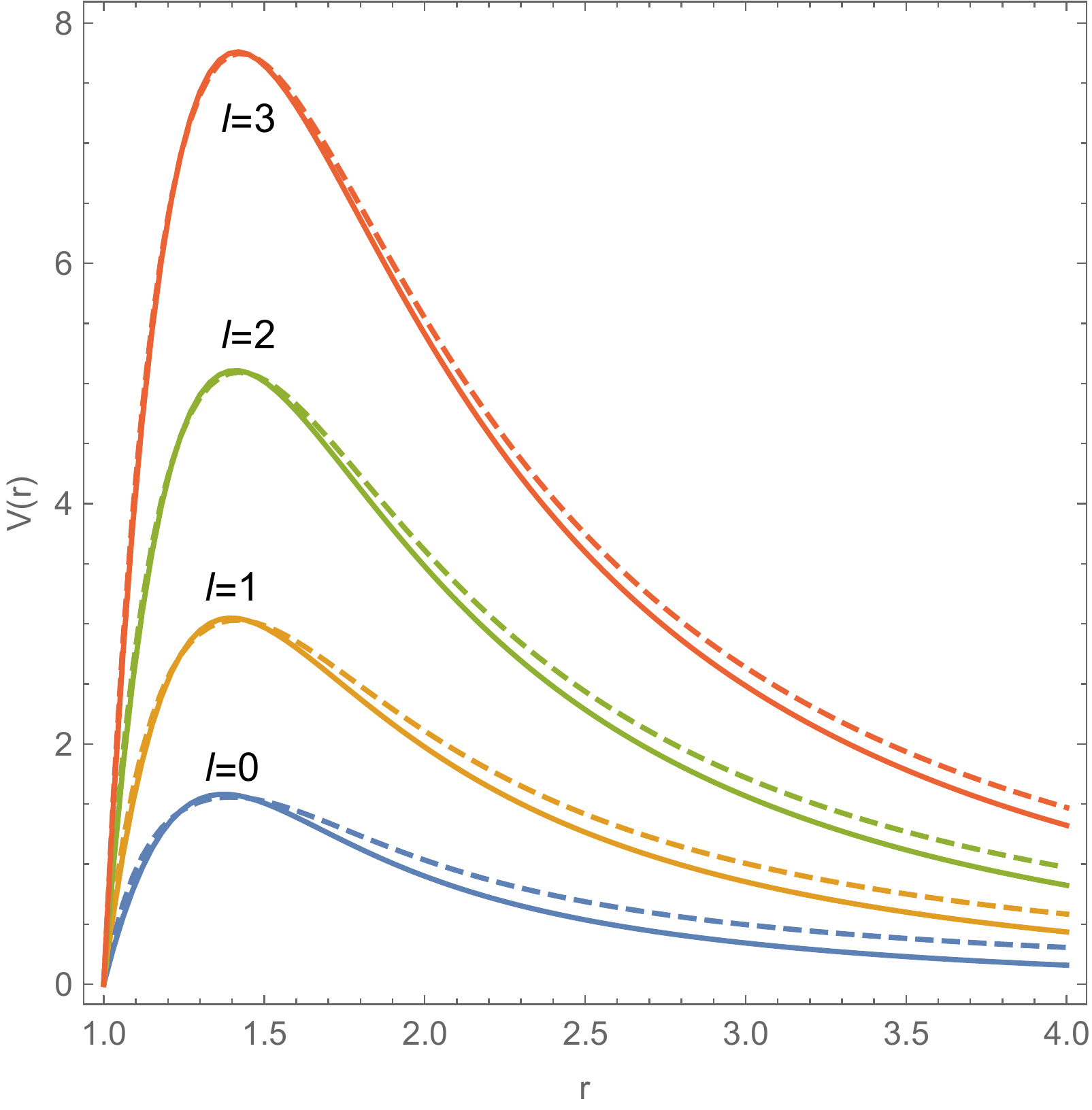}}\tabularnewline
\end{tabular}{\footnotesize\par}
\par\end{centering}
{\footnotesize{}\caption{\label{fig:XiLV} Effective potential of the scalar with the corresponding
parameters in Fig . \ref{fig:XiL}.}
}{\footnotesize\par}
\end{figure}

As we mentioned before the nonminimally coupling $\xi$ suppresses the greybody factor while the GB parameter $\tilde{\alpha}$ enhances it, thus there must be a
competition between them.
From Fig. \ref{fig:XiL}, we see that when $\tilde{\alpha}$ is small
(left panel), $\xi$ decreases the greybody factor for all modes $l$
in the entire energy regime. The situation becomes involved when $\tilde{\alpha}$
is large (right panel). In the low energy regime, $\xi$ still decreases
the greybody factor. However, it enhances the greybody factor in
the high energy regime. This behavior exists for all $l$ in $d\geq5$
EGB dS black hole spacetime. This phenomenon has an intuitive explanation
from the effective potential. From Fig. \ref{fig:XiLV}, we find that
when $\tilde{\alpha}$ is small, $\xi$ increases the effective potential
and hence suppresses the greybody factor. When $\tilde{\alpha}$ is
large, $\xi$ decreases the effective potential and enhances the greybody
factor. This is different from the Schwarzschild-de Sitter black hole
where $\xi$ always suppresses the greybody factor \cite{Pappas2016}.

{\footnotesize{}}

\subsection{\label{subsec:dgreybody}Effects of $d$ on the greybody factor }
\begin{figure}[h]
\begin{centering}
{\footnotesize{}}%
\begin{tabular}{cc}
{\footnotesize{}\includegraphics[scale=0.3]{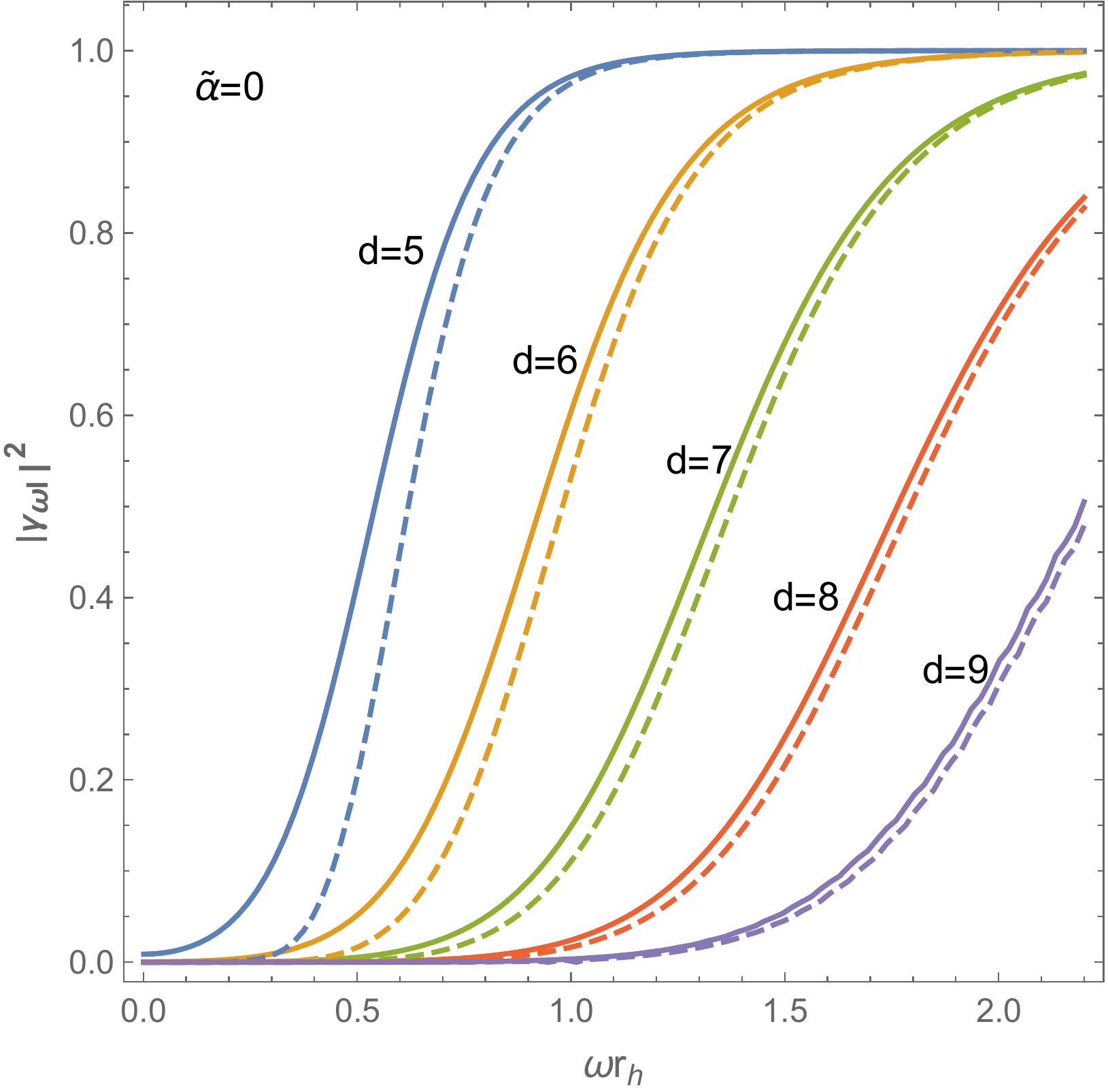}} & {\footnotesize{}\includegraphics[scale=0.3]{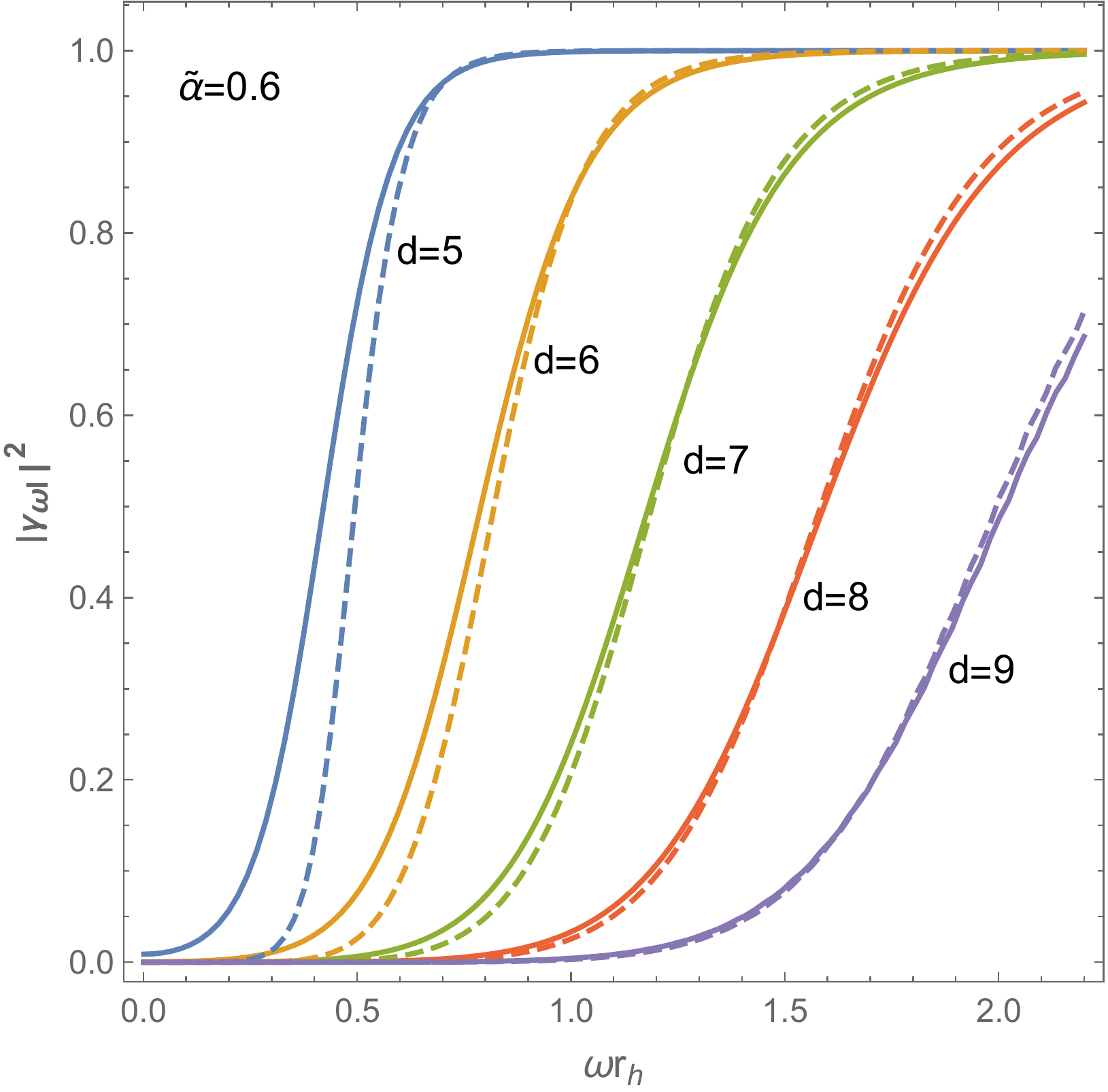}}\tabularnewline
\end{tabular}{\footnotesize\par}
\par\end{centering}
{\footnotesize{}\caption{\label{fig:AlphaD} Effect of $d$ on the greybody factor. Solid
lines for $\xi=0,$ dashed lines for $\xi=0.6$. We fix $l=0,\Lambda=0.1$
here. {The small wiggles for $d=9$ is caused by the numerical error.}}
}{\footnotesize\par}
\end{figure}

Now let us study the dependence of the greybody factor
on the spacetime dimensions $d$ in the presence of $\tilde{\alpha}$.
In Fig. \ref{fig:AlphaD}, we depict the effects of the dimensions $d$ on the greybody
factor for dominant mode $l=0$. Note that we can obtain the greybody
factors for spacetime with odd dimension $d$ here which is absent
in the approximatively analytical method in \cite{Zhang2017} due to the
poles of the Gamma function. It is obvious that the greybody factor
is significantly suppressed in higher dimensions. Furthermore, we find
similar behaviors as in subsection \ref{subsec:XiL}: when $\tilde{\alpha}$
is small, the scalar coupling constant $\xi$ always decreases the
greybody factor. When $\tilde{\alpha}$ is large, $\xi$ decreases
the greybody factor only in the low energy regime but increases the greybody factor in  the high energy
region. The behaviors of the effective
potential are also similar with those in Fig. \ref{fig:XiLV}.

{\footnotesize{}}

Note that when $\omega\to0$, the greybody factor is nonzero for dominant
mode $l=0$ for minimally coupled scalar. For $d=5,6,7,8,9$ as examples,
when $\Lambda=0.1$, the greybody factors have values of order $10^{-3},10^{-4},10^{-5},10^{-7},10^{-8}$,
respectively.

\subsection{Competition between $\Lambda$ and $\xi$ in the presence of $\tilde{\alpha}$ }

\begin{figure}[h]
\begin{centering}
{\footnotesize{}}%
\begin{tabular}{ccc}
{\footnotesize{}\includegraphics[scale=0.3]{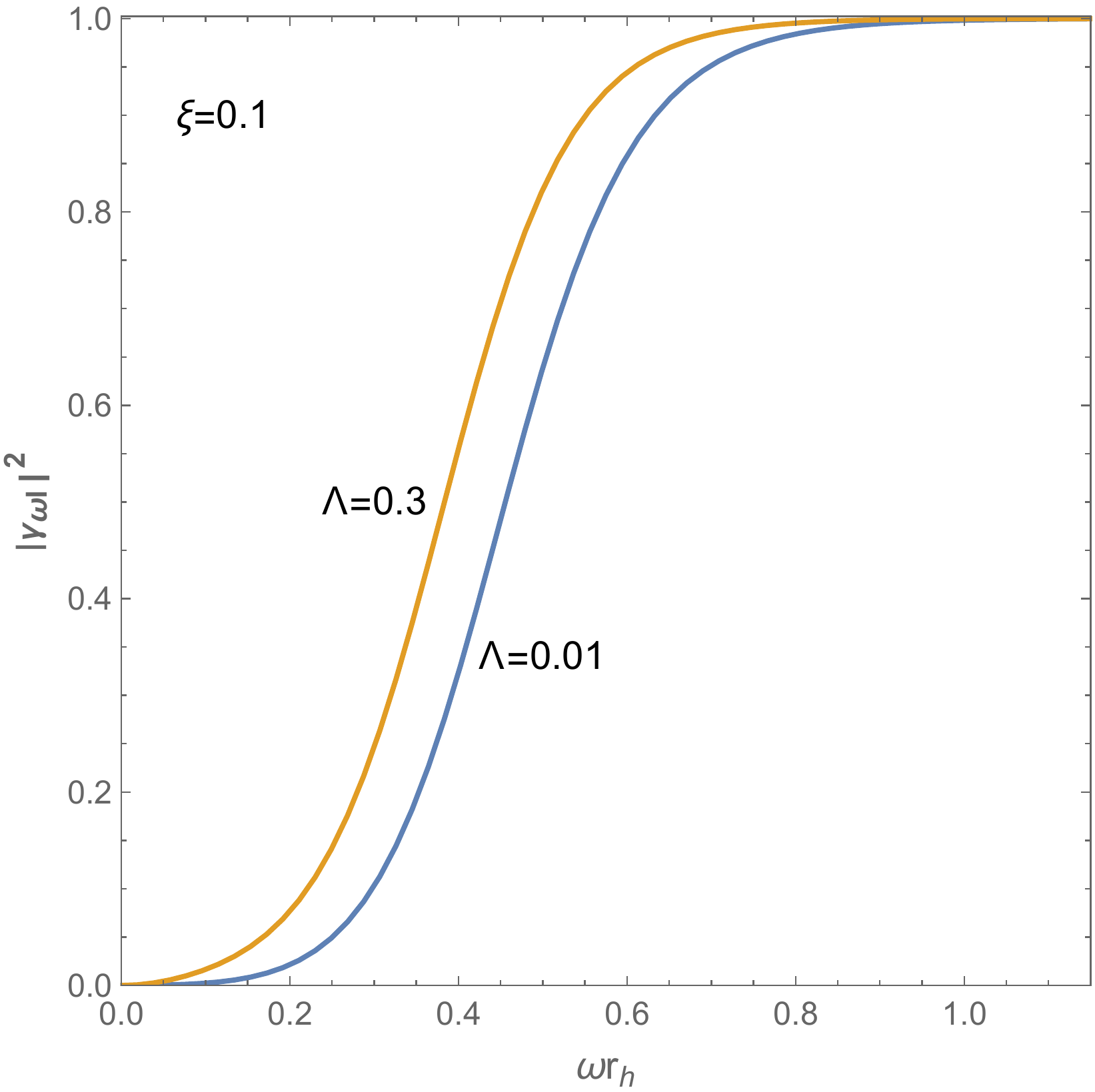}} & {\footnotesize{}\includegraphics[scale=0.3]{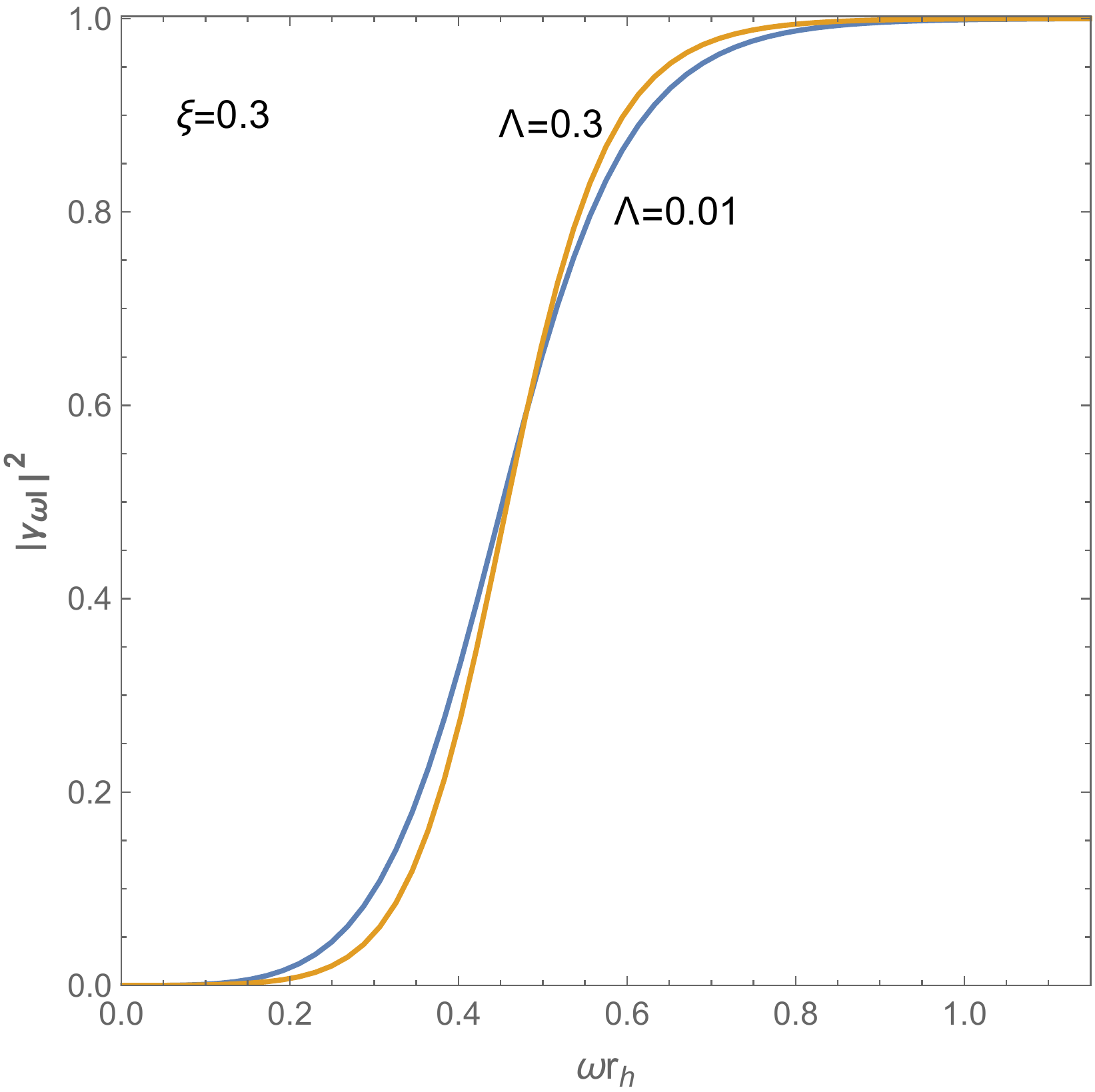}} & {\footnotesize{}\includegraphics[scale=0.3]{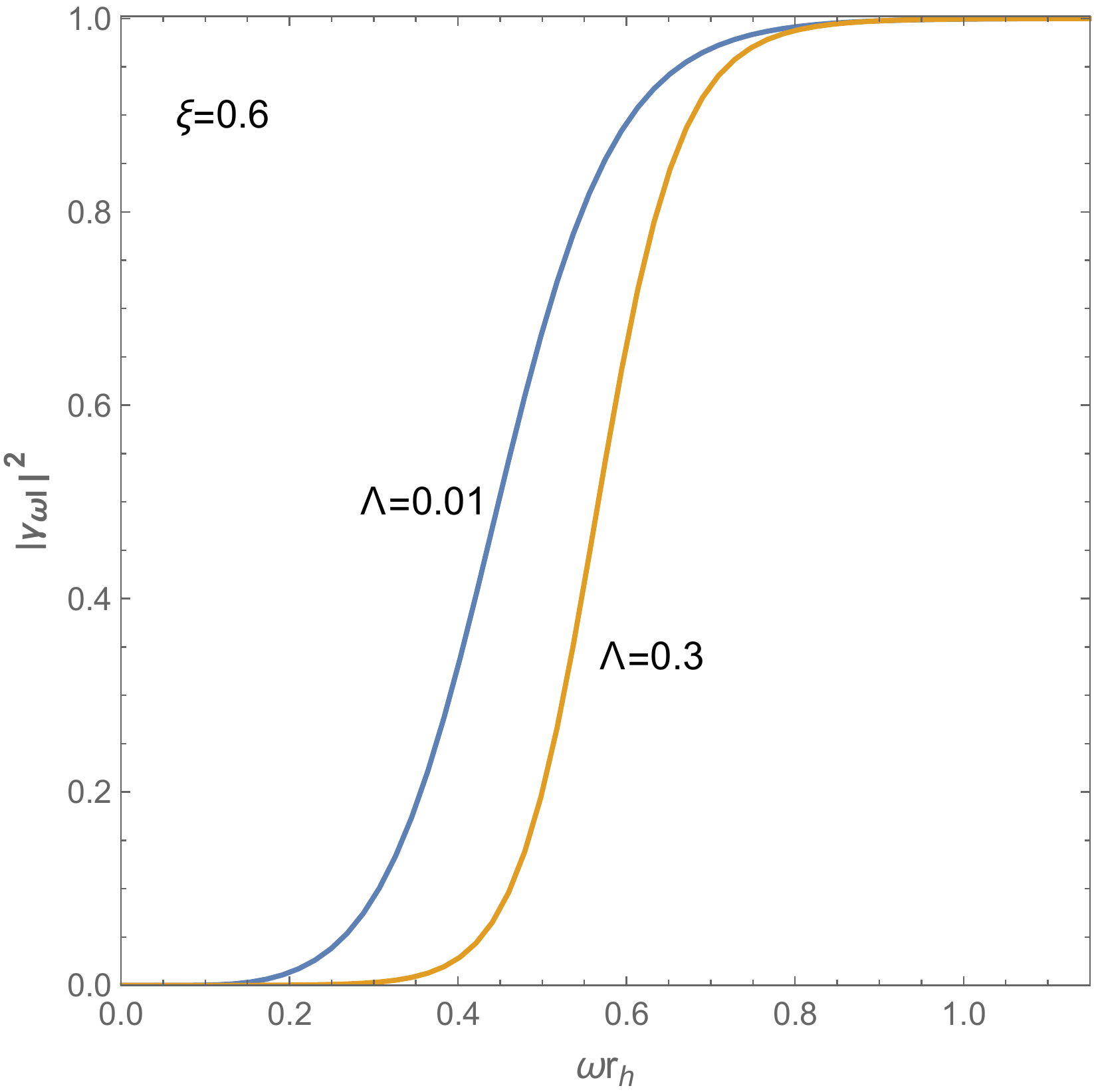}}\tabularnewline
\end{tabular}{\footnotesize\par}
\par\end{centering}
{\footnotesize{}\caption{\label{fig:AlphaXi} Competition between $\Lambda$ and $\xi$ in the presence of $\tilde{\alpha}$. The greybody factors for
 $d=5,l=0,\tilde{\alpha}=0.6$ with respect to $\Lambda=0.01, 0.3$ and $\xi=0.1, 0.3, 0.6$. }
}{\footnotesize\par}
\end{figure}

As can be seen from Fig. \ref{fig:AlphaXi}, $\Lambda$ increases
the greybody factor when $\xi$ is small in the whole energy regime
(left panel). When $\xi$ is large (right panel), $\Lambda$ decreases
the greybody factor, mainly in the low and intermediate energy regime.
This behavior has been found in \cite{Zhang2017} by using approximate
analytical method which is reliable only in low energy regime there.
{This phenomenon can be understood from the double roles $\Lambda$ plays.
The positive cosmological constant behaves like an homogenously distributed energy in the spacetime. It increases the effective energy of the particles, making it easier to pass through the effective potential barrier. Hence it can enhance the greybody factor. On the other hand, it couples to $\xi$ in the equation of motion (\ref{eq:EOMs}) and behaves as an effective mass for the particles, making it harder to transverse the effective potential. Hence it can also suppress the greybody factor.
}
%
For small $\xi$, the suppression due to the effective mass is also
small, as a consequence the enhancement effect of cosmological constant dominates the
process. For large $\xi$, the effective mass increases substantially
with $\Lambda$, thus the greybody factor is suppressed.
{This phenomenon is also discovered in other dimensions.}
We find that this competition behavior of $\Lambda$ and $\xi$ is independent of
the GB coupling $\tilde{\alpha}$.

{Similarly, the effect of $\tilde{\alpha}$ on the greybody factor as analyzed in \ref{subsec:AlphaL} could be explained from a more physical way. First, from the field equations of the theory (\ref{eq:EGBAction})
\begin{equation}
R_{\mu\nu}-\frac{1}{2}R g_{\mu\nu}+\Lambda g_{\mu\nu}+\alpha H_{\mu\nu}=0,
\end{equation}
where
\begin{equation}
H_{\mu\nu}=-\frac{1}{2}g_{\mu\nu}\mathcal{L}_{GB}+2(RR_{\mu\nu}-2R_{\mu\gamma}R^{\gamma}_{\,\,\nu}+2R^{\gamma\delta}R_{\gamma\mu\nu\delta}
+R_{\mu\gamma\delta\lambda}R_{\nu}^{\,\,\gamma\delta\lambda}),
\end{equation}
formally, one may treat $-\alpha H_{\mu\nu}$ as an effective stress tensor. Then from this point of view, the quantity $\alpha H_{tt}$ can be regarded as a local energy density. It is straightforward to check that this quantity is always positive for a positive $\alpha$, which as a result could increase the effective energy of the particles and enhance its ability of  tunneling through the effective potential barrier.
 }

{\footnotesize{}}

\section{Energy emission rate of Hawking radiation \label{sec:Energy-emission-rate}}

The greybody factor reflects only the transmissity of a particular
mode. A more comprehensive quantity that characterizes the Hawking
radiation is the energy emission rate, i.e., the power spectra for
the emission. The differential energy
emission rate is given by \cite{Kanti2004,Harris2003,Kanti2005}
\begin{equation}
\frac{d^{2}E}{dtd\omega}=\frac{1}{2\pi}\sum_{l}\frac{N_{l}|\gamma_{\omega l}|^{2}\omega}{e^{\omega/T}-1},\label{eq:PowerSpectra}
\end{equation}
 where $N_{l}=\frac{(2l+d-3)(l+d-4)!}{l!(d-3)!}$ is the multiplicity
of states that have the same $l$.
$T$ is the temperature of the system. For black hole spacetime, it
is usually defined as $T_{0}=\frac{\kappa_{h}}{2\pi}$, where $\kappa_{h}$
is the surface gravity of the event horizon. However, this definition
is valid only in asymptotically flat spacetime \cite{Bousso1996}. For
asymptotic dS spacetime, various definitions of temperature were proposed.
Here we will consider six different temperatures and study their influences
on the power spectra of Hawking radiation. The normalized temperature
that proposed in \cite{Bousso1996} is
\begin{equation}
T_{BH}=\frac{1}{\sqrt{h(r_{0})}}\frac{\kappa_{h}}{2\pi},
\end{equation}
were $r_{0}$ is the position where $h(r)$ is extreme.  The black hole
attraction balances the cosmological repulsion near this point such
that the spacetime is almost flat there. For dS black hole, there
is cosmological horizon temperature defined as $T_{c}=-\frac{\kappa_{c}}{2\pi},$
where $\kappa_{c}$ is the surface gravity of the cosmological horizon.
In general, $T_{0}$ and $T_{c}$ are different, thus the spacetime
is not in equilibrium. Some effective temperatures inspired from the
black hole thermodynamics were proposed to resolve this problem (See
\cite{RPappasb,Pappas2016,Bousso1996,Kanti2017} for more information):
\begin{align}
T_{effEIW}=\frac{r_{h}^{4}T_{c}+r_{c}^{4}T_{0}}{(r_{h}+r_{c})(r_{c}^{3}-r_{h}^{3})}, & \quad T_{effBH}=\left(\frac{1}{T_{c}}-\frac{1}{T_{BH}}\right)^{-1},\\
T_{eff-}=\left(\frac{1}{T_{c}}-\frac{1}{T_{0}}\right)^{-1},\quad & T_{eff+}=\left(\frac{1}{T_{c}}+\frac{1}{T_{0}}\right)^{-1}.
\end{align}

\begin{figure}[h]
\begin{centering}
{\footnotesize{}}%
\begin{tabular}{ccc}
{\footnotesize{}\includegraphics[scale=0.28]{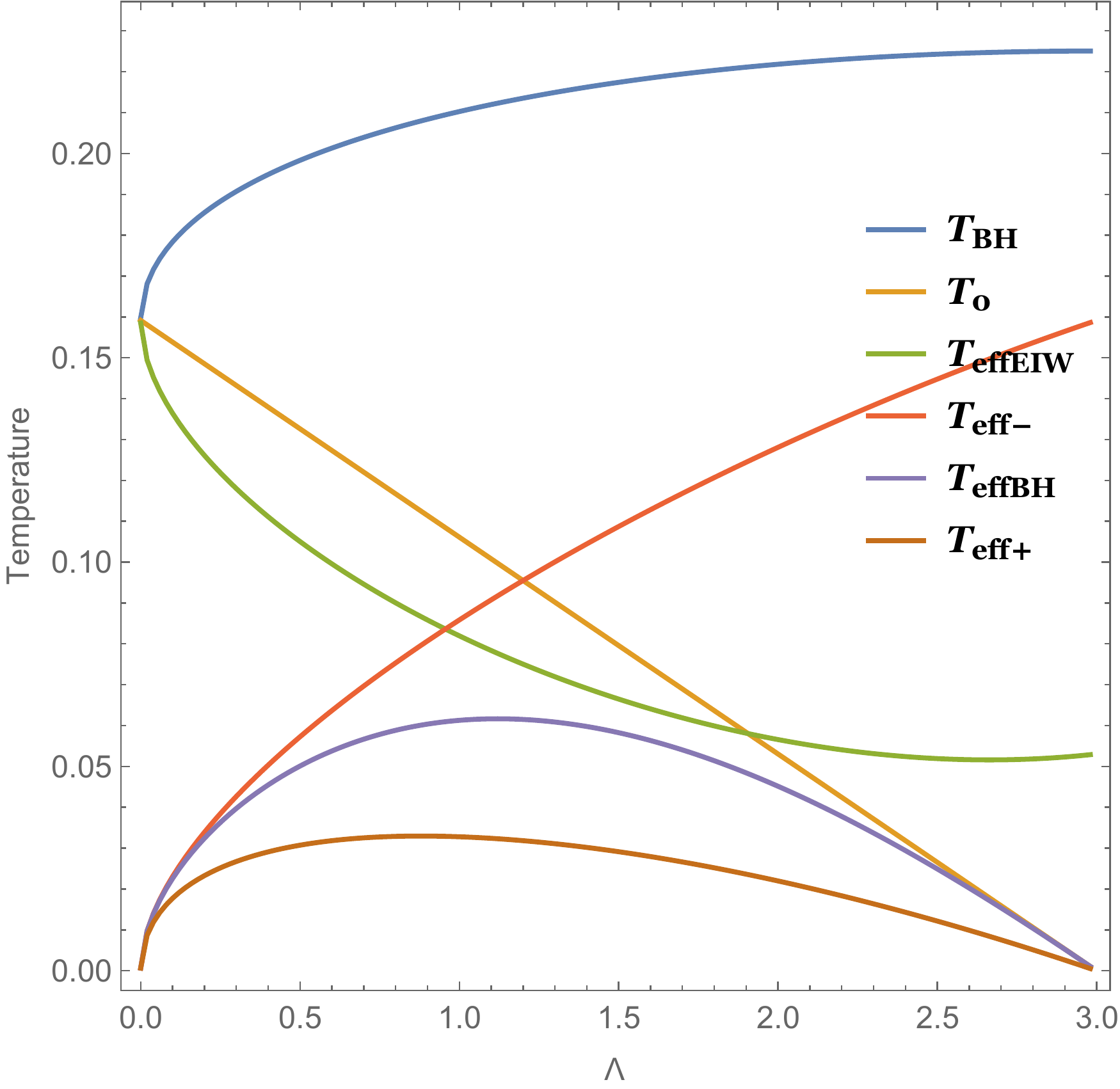}} & {\footnotesize{}\includegraphics[scale=0.28]{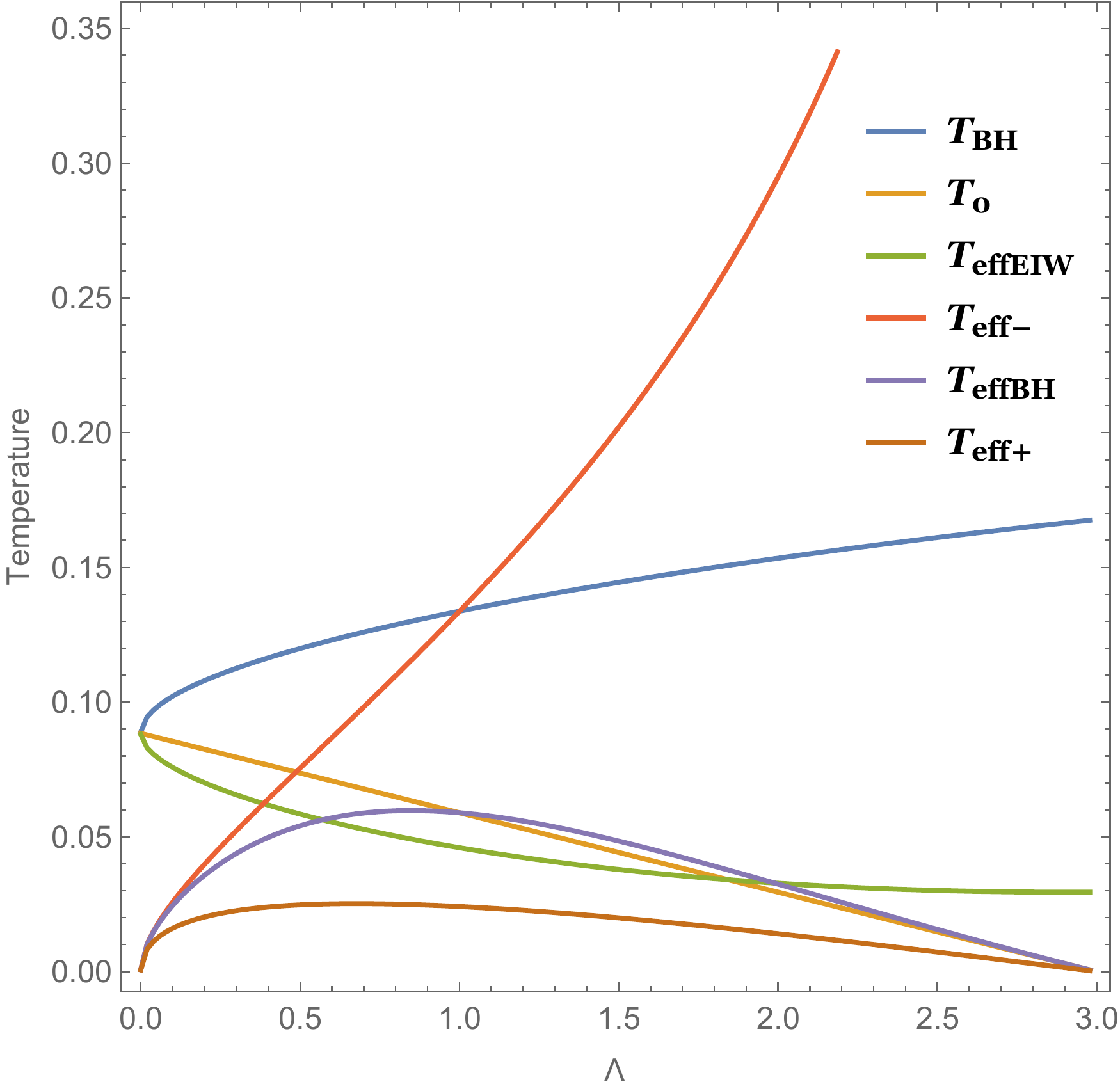}}& {\footnotesize{}\includegraphics[scale=0.3]{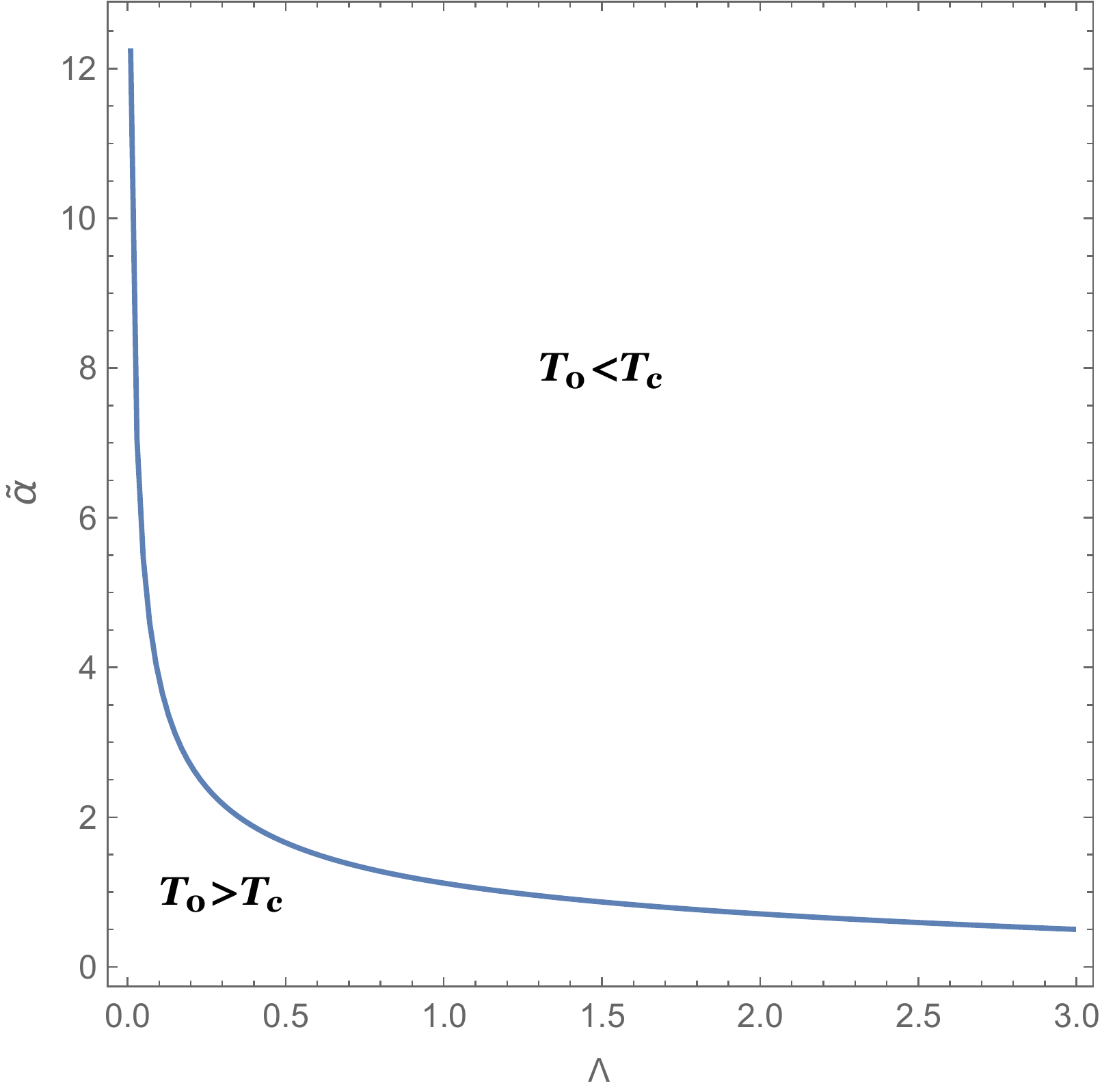}}\tabularnewline
\end{tabular}{\footnotesize\par}
\par\end{centering}
{\footnotesize{}\caption{\label{fig:TempL} Different temperatures with respect to $\Lambda$ when $d=5$. The extremal cosmological constant is $\Lambda=3$ where $r_c=r_h$.
Left panel for $\tilde{\alpha}=0$, middle panel for $\tilde{\alpha}=0.4$, right panel for the critical curve where $T_0=T_c$. In higher dimensions, there is always $T_0>T_c$.}
}{\footnotesize\par}
\end{figure}

{We show their dependence on $\Lambda$ in
Fig. \ref{fig:TempL} when $d=5$.  When $\tilde{\alpha}$ is small, $T_{BH}$ dominates in the whole $\Lambda$ region. If $\tilde{\alpha}$ is large, $T_{eff-}$ becomes dominant for large $\Lambda$. However, $T_{eff-}$ jumps to be negative when $\tilde{\alpha}$ exceeds a critical value where $T_0$ becomes smaller than $T_c$ in the parameter region shown in the right panel of Fig. \ref{fig:TempL}.
The greater $\Lambda$ is, the smaller critical $\tilde{\alpha}$. When $d>5$, it can be shown that there is  always $T_0 > T_c$ such that $T_{eff-}$ is always positive.
Moreover, we find that when $\Lambda \to 0$, $T_{eff-}, T_{effBH},T_{eff+}$ all tend to 0. This will lead to vanishing radiation in asymptotically flat spacetime and is unreasonable. Thus we will take $T_{BH},T_{effEIW}$ as the temperature in (\ref{eq:PowerSpectra}).}

\begin{figure}[h]
\begin{centering}
{\footnotesize{}}%
\begin{tabular}{cc}
{\footnotesize{}\includegraphics[scale=0.3]{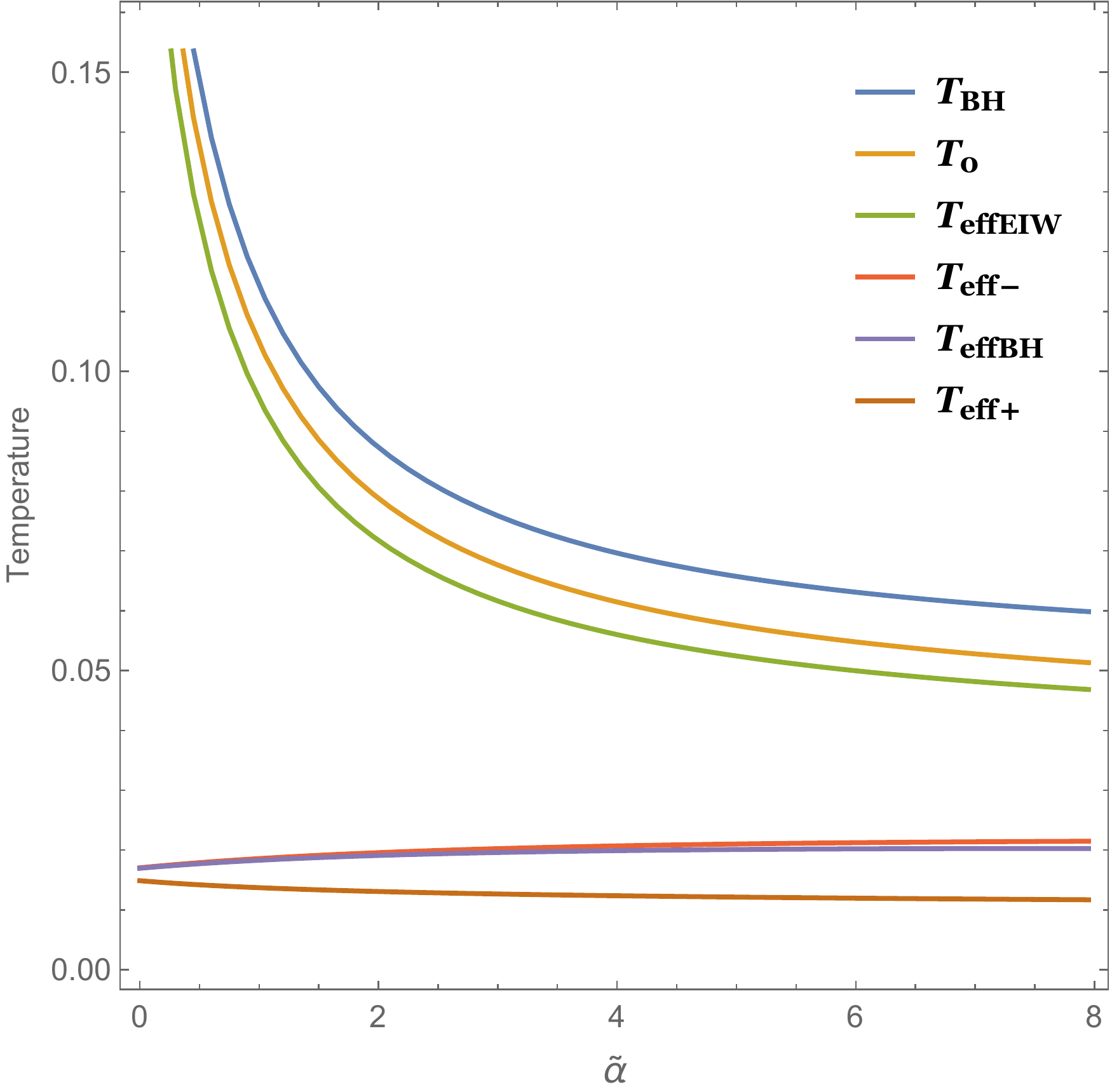}} & {\footnotesize{}\includegraphics[scale=0.3]{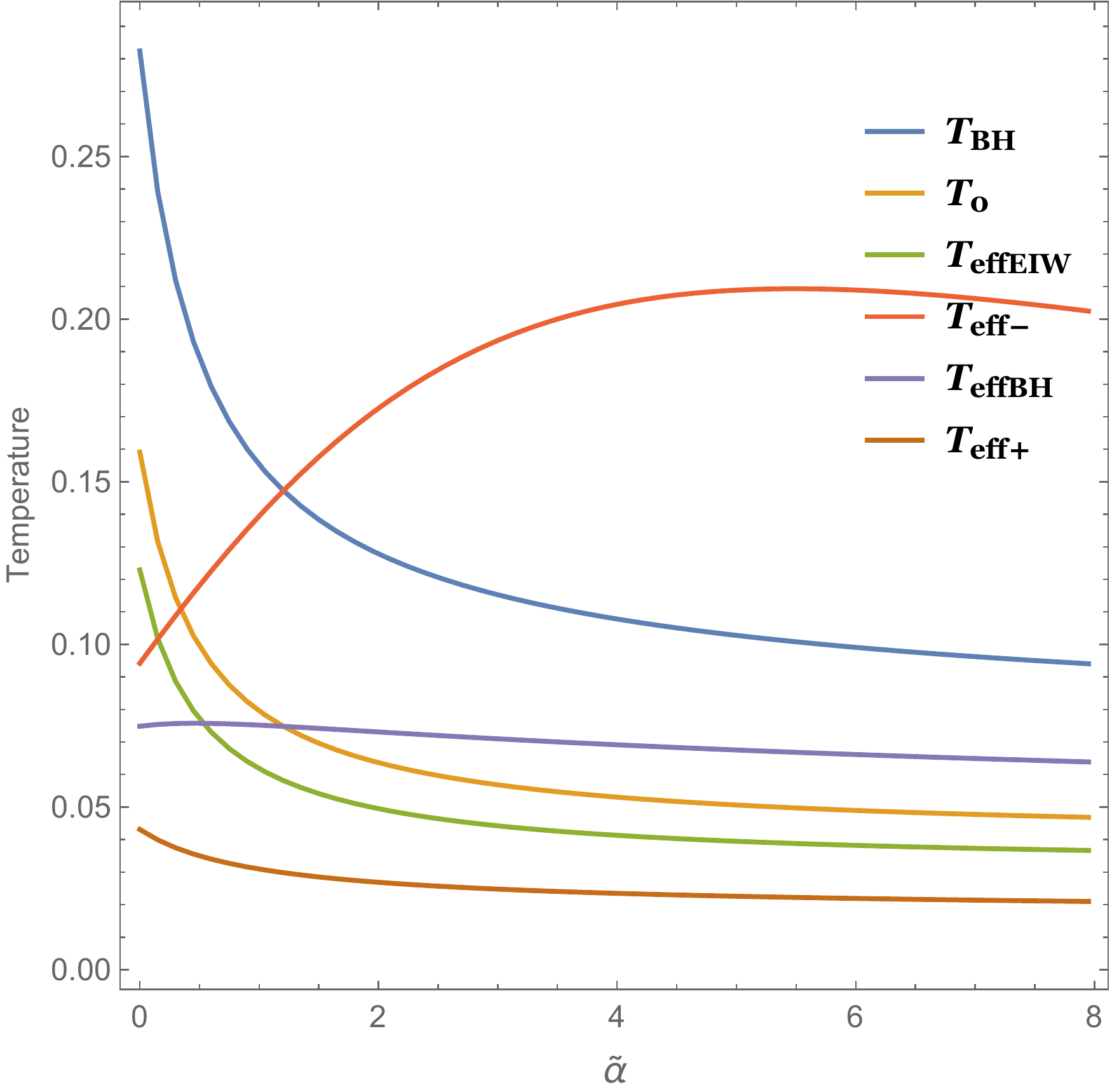}}\tabularnewline
\end{tabular}{\footnotesize\par}
\par\end{centering}
{\footnotesize{}\caption{\label{fig:TempA} Different temperatures with respect to $\tilde{\alpha}$ when  $d=6$. The extremal cosmological constant is $\Lambda=6+2\tilde{\alpha}$.
Left panel for $\Lambda=0.1$, right panel for $\Lambda=1$. }
}{\footnotesize\par}
\end{figure}

{Figure \ref{fig:TempA} shows the influence of $\tilde{\alpha}$ on temperature when $d=6$. We find that $T_{BH},T_{effEIW}$ all decrease with  $\tilde{\alpha}$ no matter whether $\Lambda$ is small or large. This behavior is also observed in other dimensions. This enables us to take $T_{BH}$ as example to study the power spectra of Hawking radiation. The qualitative behavior of power spectra with respect to $\tilde{\alpha}$ will not change if $T_0$ or $T_{effEIW}$ is adopted.
In the following, we will only show the results of $d=5$ in most part of this section for convenience. Qualitatively similar behaviors are observed in higher dimensions.}

{\footnotesize{}}

{\footnotesize{}}

\begin{figure}[h]
\begin{centering}
{\footnotesize{}}%
\begin{tabular}{cc}
{\footnotesize{}\includegraphics[scale=0.3]{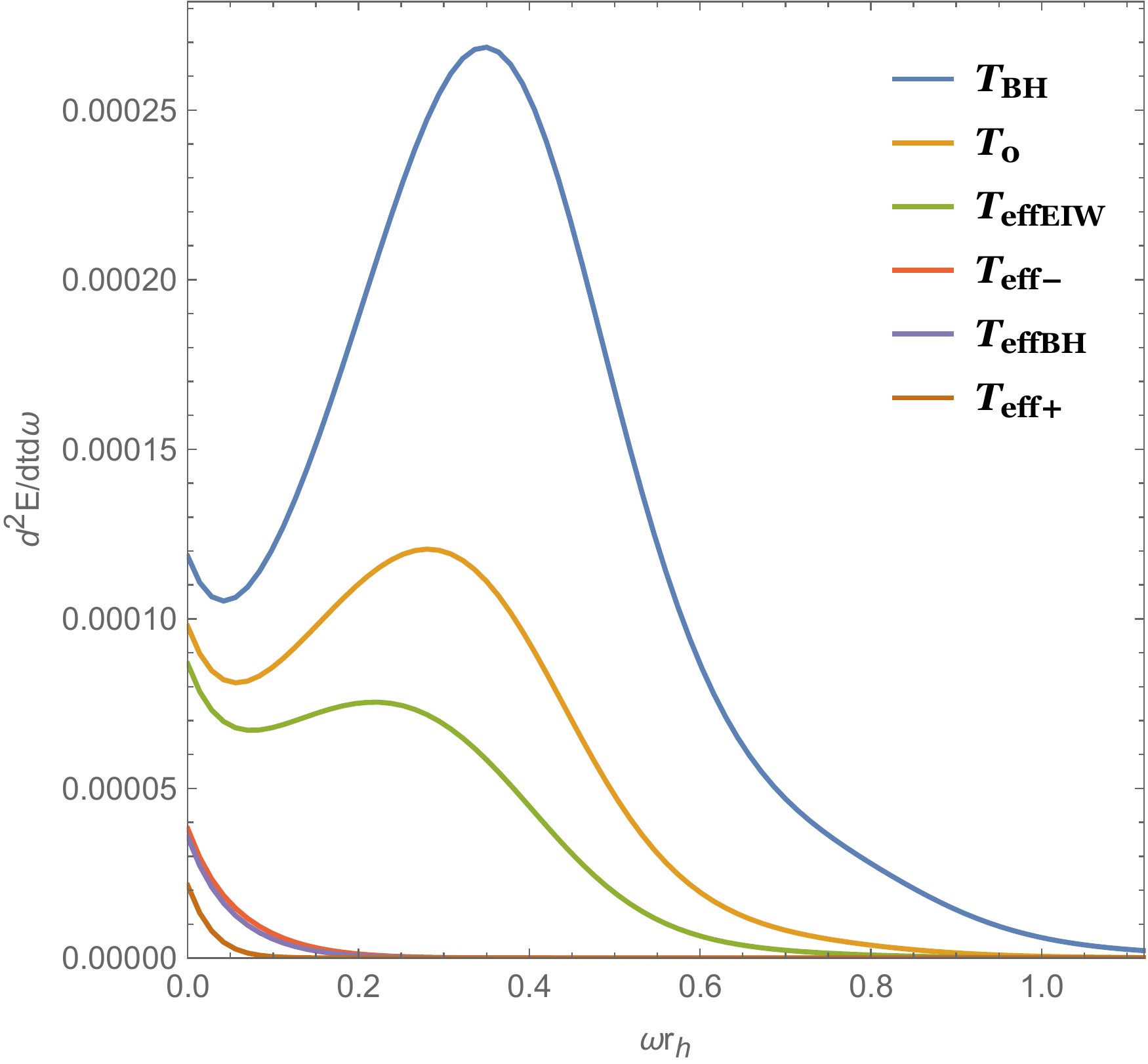}} & {\footnotesize{}\includegraphics[scale=0.3]{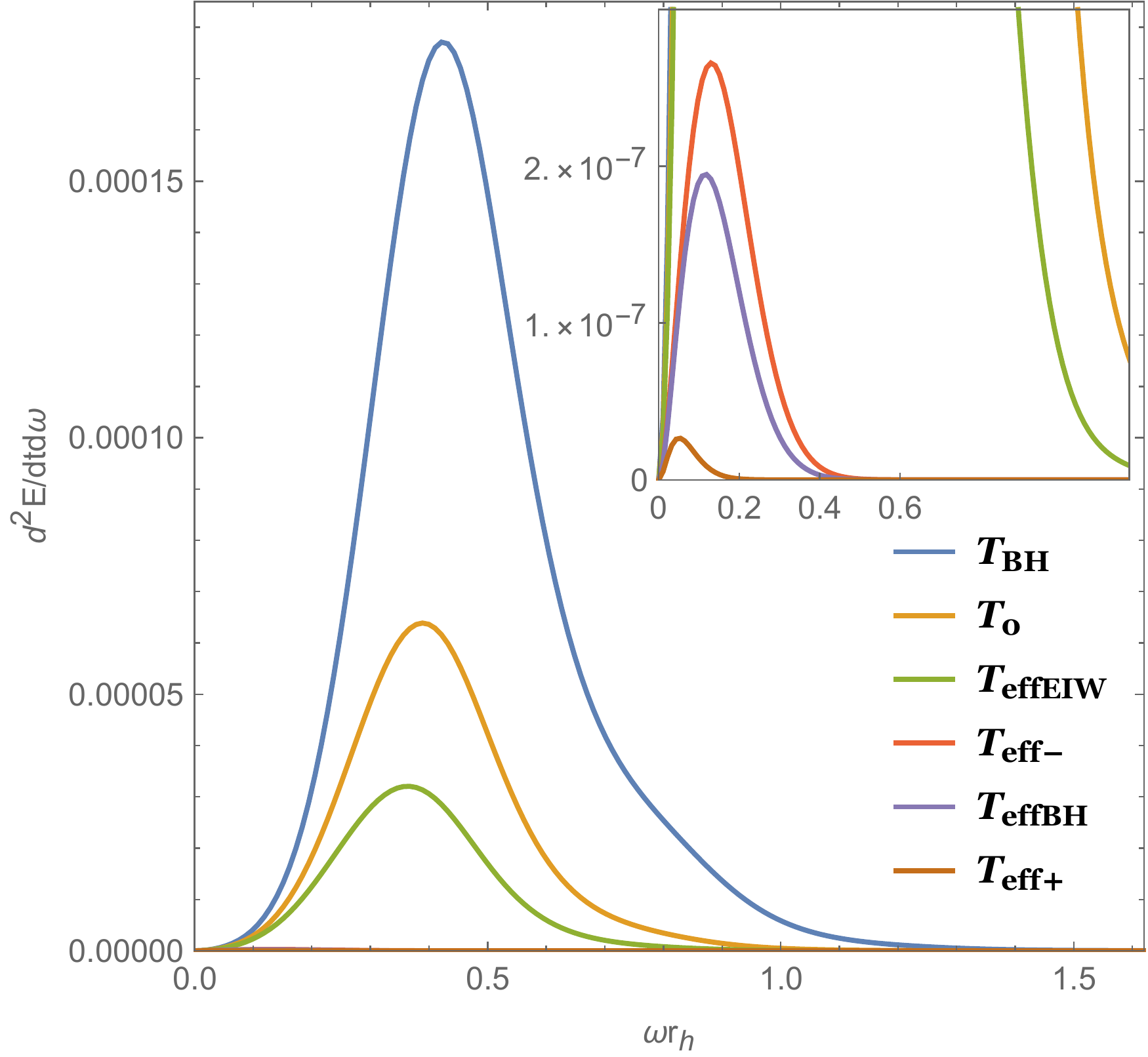}}\tabularnewline
\end{tabular}{\footnotesize\par}
\par\end{centering}
{\footnotesize{}\caption{\label{fig:TempXiLambdaSmall} Power spectra for different temperatures.
Left panel for $\xi=0$, right panel for $\xi=0.3$. We fix $d=5$,
$\Lambda=0.1$, $\tilde{\alpha}=0.6$ here.}
}{\footnotesize\par}
\end{figure}

\begin{figure}[h]
\begin{centering}
{\footnotesize{}}%
\begin{tabular}{cc}
{\footnotesize{}\includegraphics[scale=0.3]{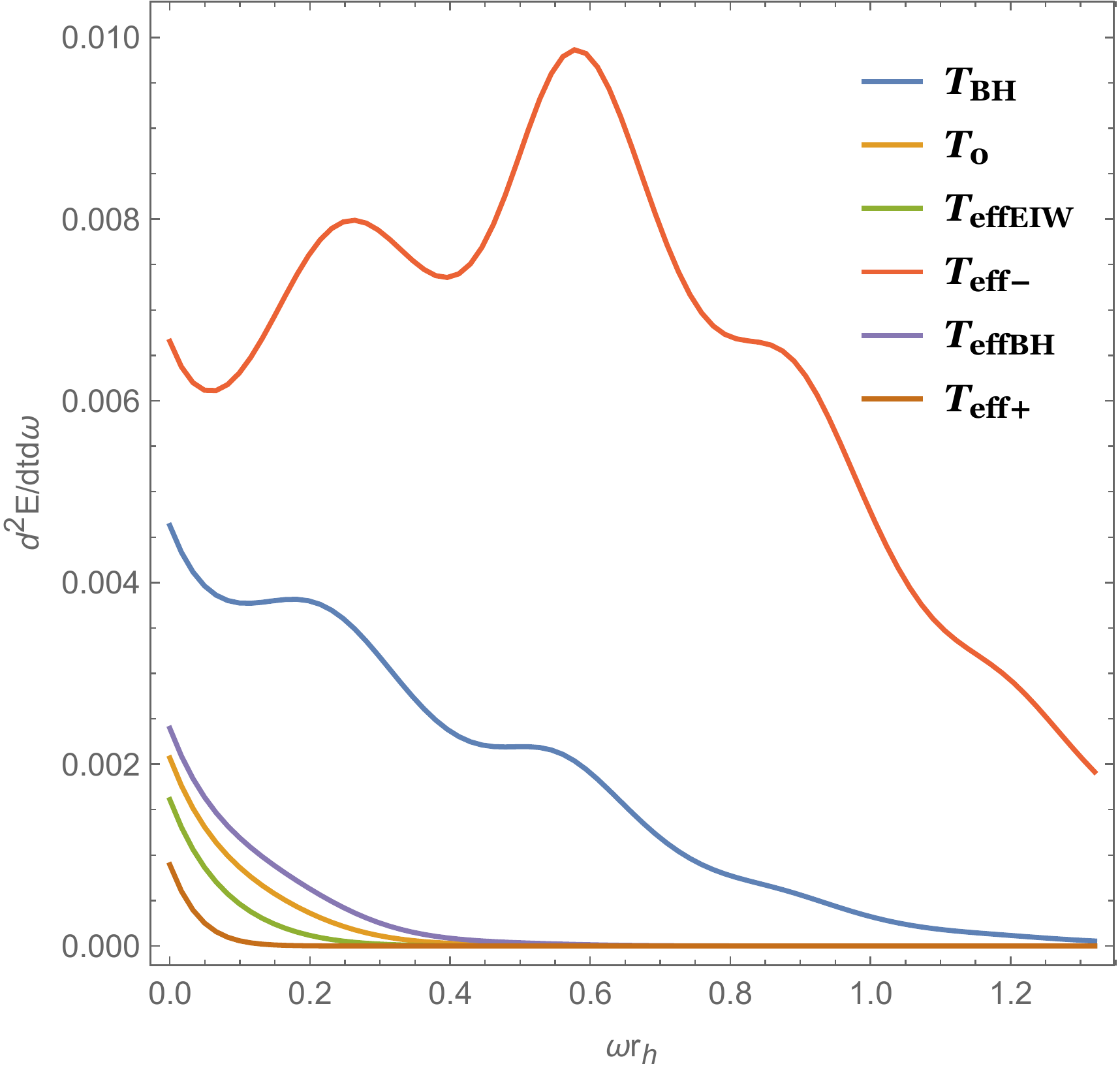}} & {\footnotesize{}\includegraphics[scale=0.3]{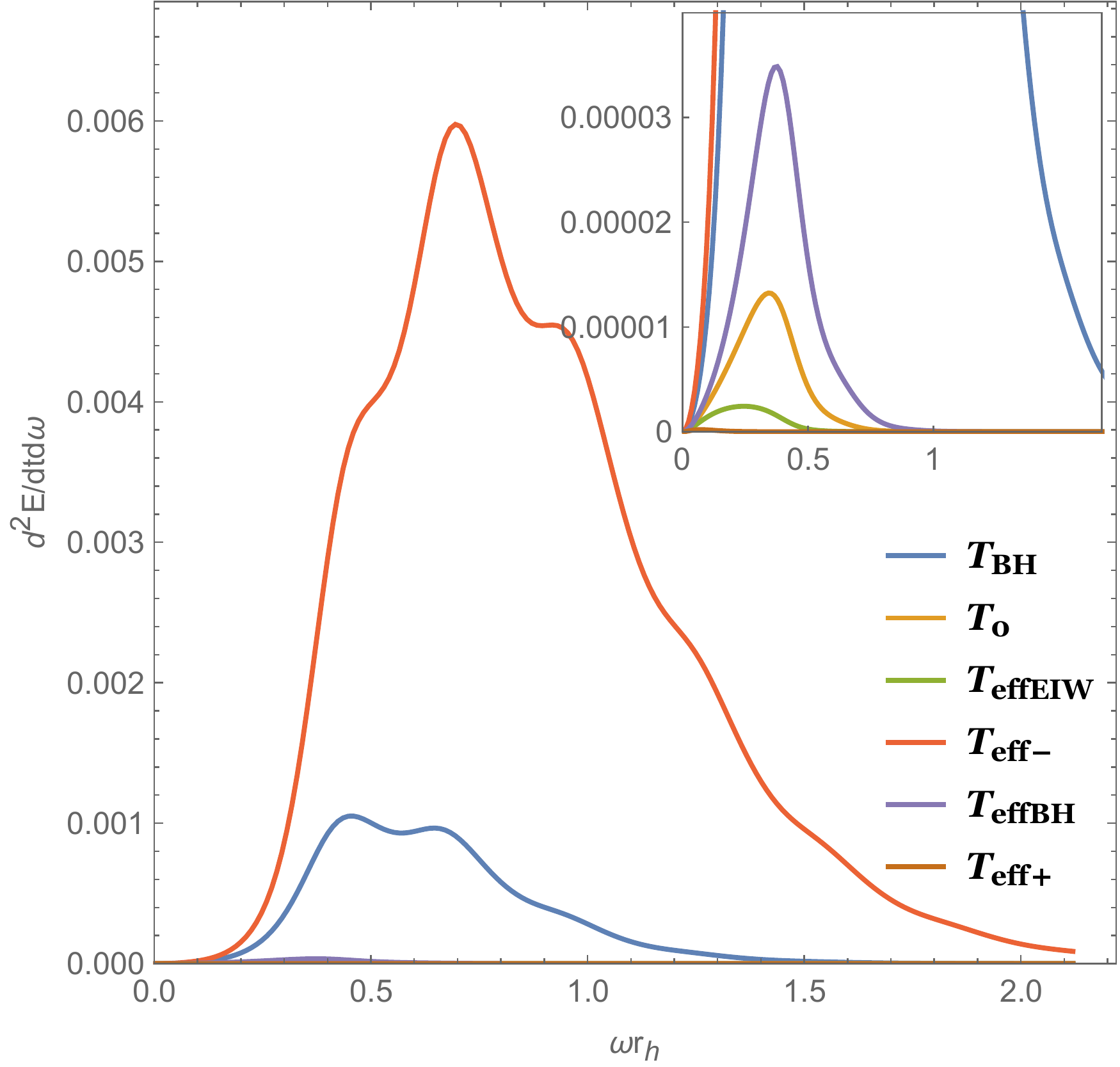}}\tabularnewline
\end{tabular}{\footnotesize\par}
\par\end{centering}
{\footnotesize{}\caption{\label{fig:TempXiLambdaLarge} Power spectra for different temperatures.
Left panel for $\xi=0$, right panel for $\xi=0.3$. We fix $d=5$,
$\Lambda=0.9$, $\tilde{\alpha}=0.6$ here.}
}{\footnotesize\par}
\end{figure}

Substituting these temperatures into (\ref{eq:PowerSpectra}), we
get the power spectra shown in Fig. \ref{fig:TempXiLambdaSmall} and
Fig. \ref{fig:TempXiLambdaLarge} for examples. For minimally coupled
scalar, the power spectra in the low energy limit is finite, while
the power spectra for nonminimally coupled scalar in the low energy
limit vanishes. This is due to the nonvanishing greybody factors
for the dominant mode $l=0$ of minimally coupled scalar, as shown
in (\ref{eq:ALowLimit}).

{\footnotesize{}}

{\footnotesize{}}

Moreover, We can see that only $T_{BH}$ leads to significant radiation
both for small $\Lambda$ and large $\Lambda$. We therefore take
this temperature to study the effects of other parameters on the Hawking
radiation hereafter.

\begin{figure}[h]
\begin{centering}
{\footnotesize{}}%
\begin{tabular}{c}
{\footnotesize{}\includegraphics[scale=0.3]{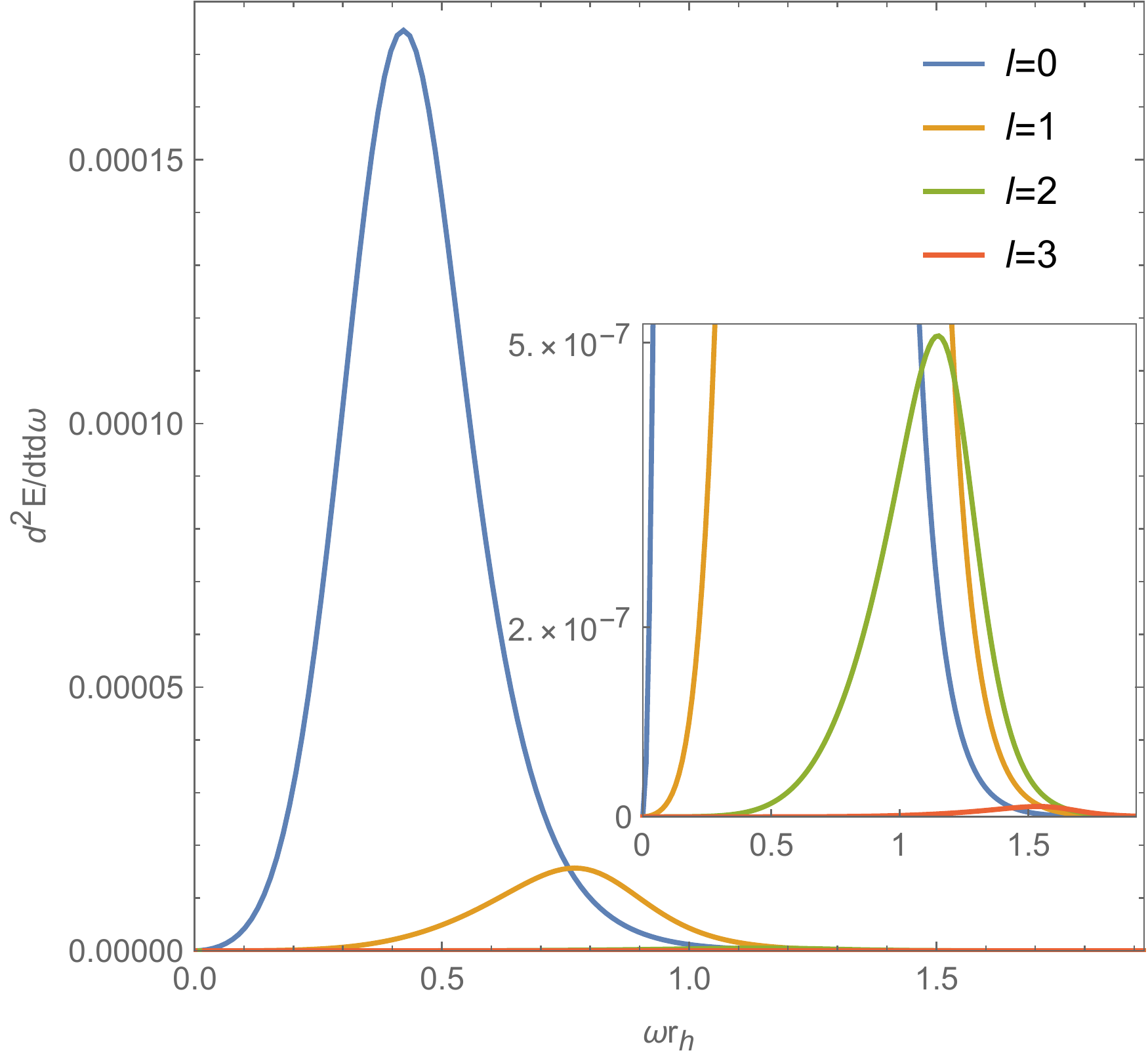}}\tabularnewline
\end{tabular}{\footnotesize\par}
\par\end{centering}
{\footnotesize{}\caption{\label{fig:PowerL} The contributions from different $l$. We fix $d=5,\xi=0.3,\tilde{\alpha}=0.3,\Lambda=0.1$ here. Note that the contribution from $l=3$ is of order $10^{-8}$.}
}{\footnotesize\par}
\end{figure}

The contribution
to the energy emission rate comes mainly from lower modes $l$ \cite{Pappas2016,Zhang2017},
{as shown in Fig. \ref{fig:PowerL}  for example.
Without loss of validness, we could omit the contributions from modes $l>6$ in (\ref{eq:PowerSpectra})
hereafter.
}
{\footnotesize{}}

\subsection{Effects of $\tilde{\alpha}$ and $\xi$ on the power spectra of Hawking
radiation}

\begin{figure}[h]
\begin{centering}
{\footnotesize{}}%
\begin{tabular}{cc}
{\footnotesize{}\includegraphics[scale=0.3]{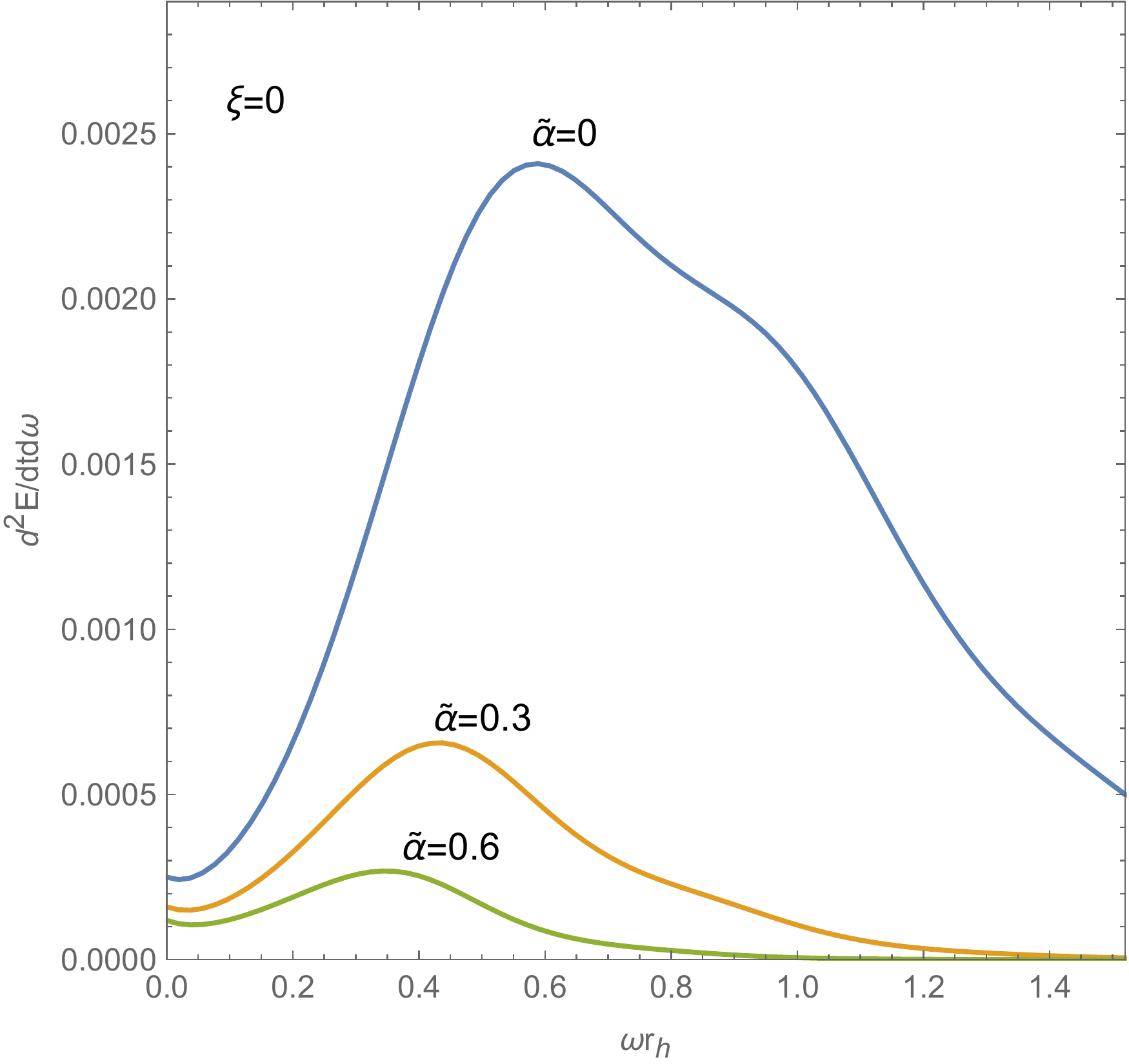}} & {\footnotesize{}\includegraphics[scale=0.3]{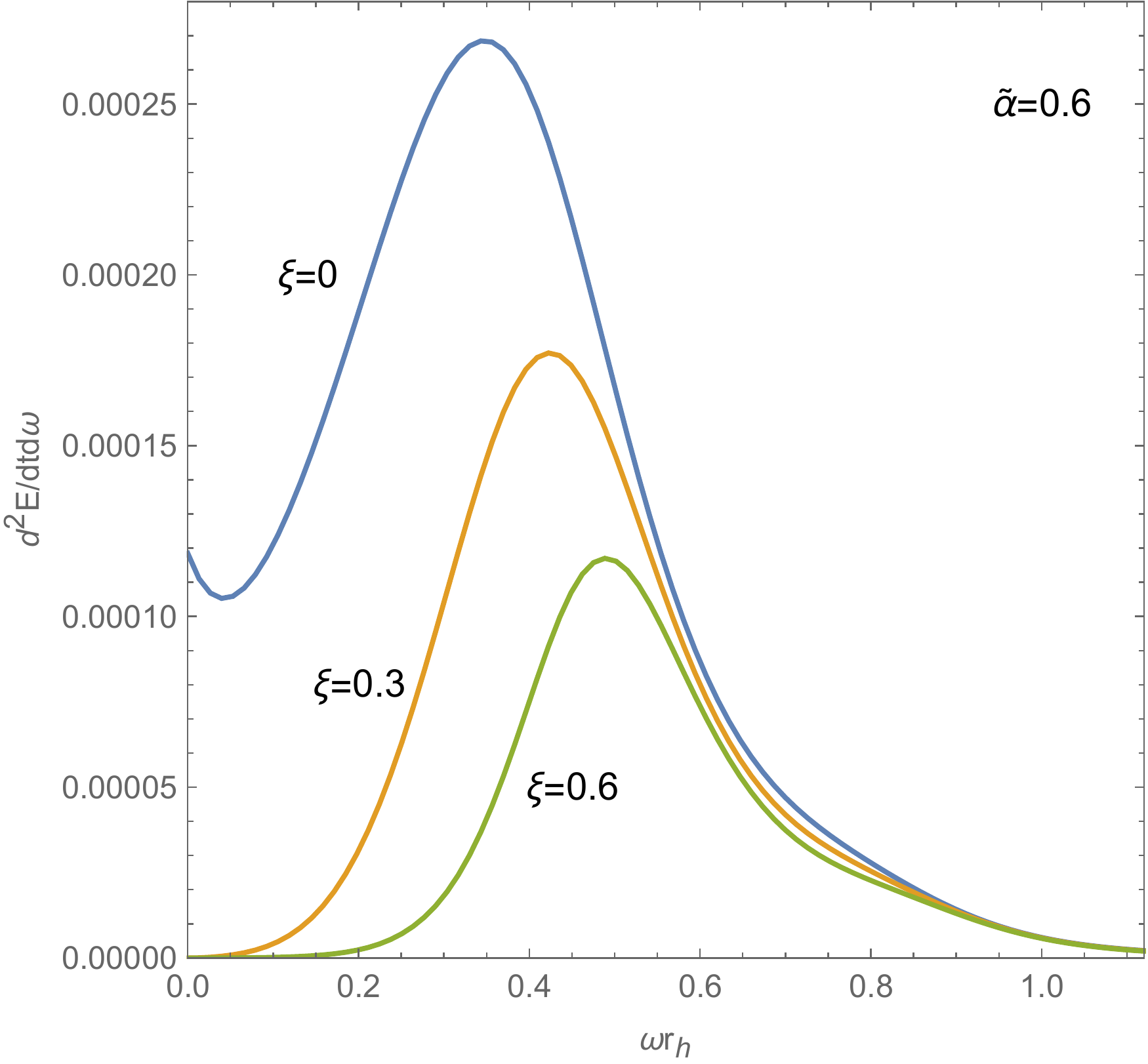}}\tabularnewline
\end{tabular}{\footnotesize\par}
\par\end{centering}
{\footnotesize{}\caption{\label{fig:PowerAlphaXi} Effect of $\tilde{\alpha}$ and $\xi$ on
the power spectra of Hawking radiation. We fix $d=5,\Lambda=0.1$
here.}
}{\footnotesize\par}
\end{figure}

We study the effects of $\tilde{\alpha}$ and $\xi$ on the power spectra
of Hawking radiation in this subsection. From the left panel of Fig.
\ref{fig:PowerAlphaXi}, we see that $\tilde{\alpha}$ suppresses
the power spectra in the whole energy regime. This may be surprised
since we have found that $\tilde{\alpha}$ decreases the greybody
factor in the entire  energy regime in subsection \ref{subsec:AlphaL}.
In fact, the power spectra depends also on the normalized temperature
$T_{BH}$. The significant suppression of the power spectra with $\tilde{\alpha}$
is due to the decrease of $T_{BH}$ as shown in Fig. \ref{fig:TempA}.

{\footnotesize{}}

On the other hand, in subsections \ref{subsec:XiL} and \ref{subsec:dgreybody},
we found that $\xi$ could increase the greybody factor in the high
energy regime when $\tilde{\alpha}$ is large enough. However, we
find that $\xi$ always suppresses the power spectra in the whole
energy regime, as shown in the right panel of Fig. \ref{fig:PowerAlphaXi}.
We find that this qualitative behavior is independent of $\tilde{\alpha}$.
Moreover, when other parameters are fixed, the peak of the power spectra
moves to lower frequency when $\tilde{\alpha}$ increases, or to the
higher frequency when $\xi$ increases. These behaviors are in accordance
with those found in \cite{Zhang2017} by using approximative analytical
method.

\begin{figure}[h]
\begin{centering}
{\footnotesize{}}%
\begin{tabular}{c}
{\footnotesize{}\includegraphics[scale=0.35]{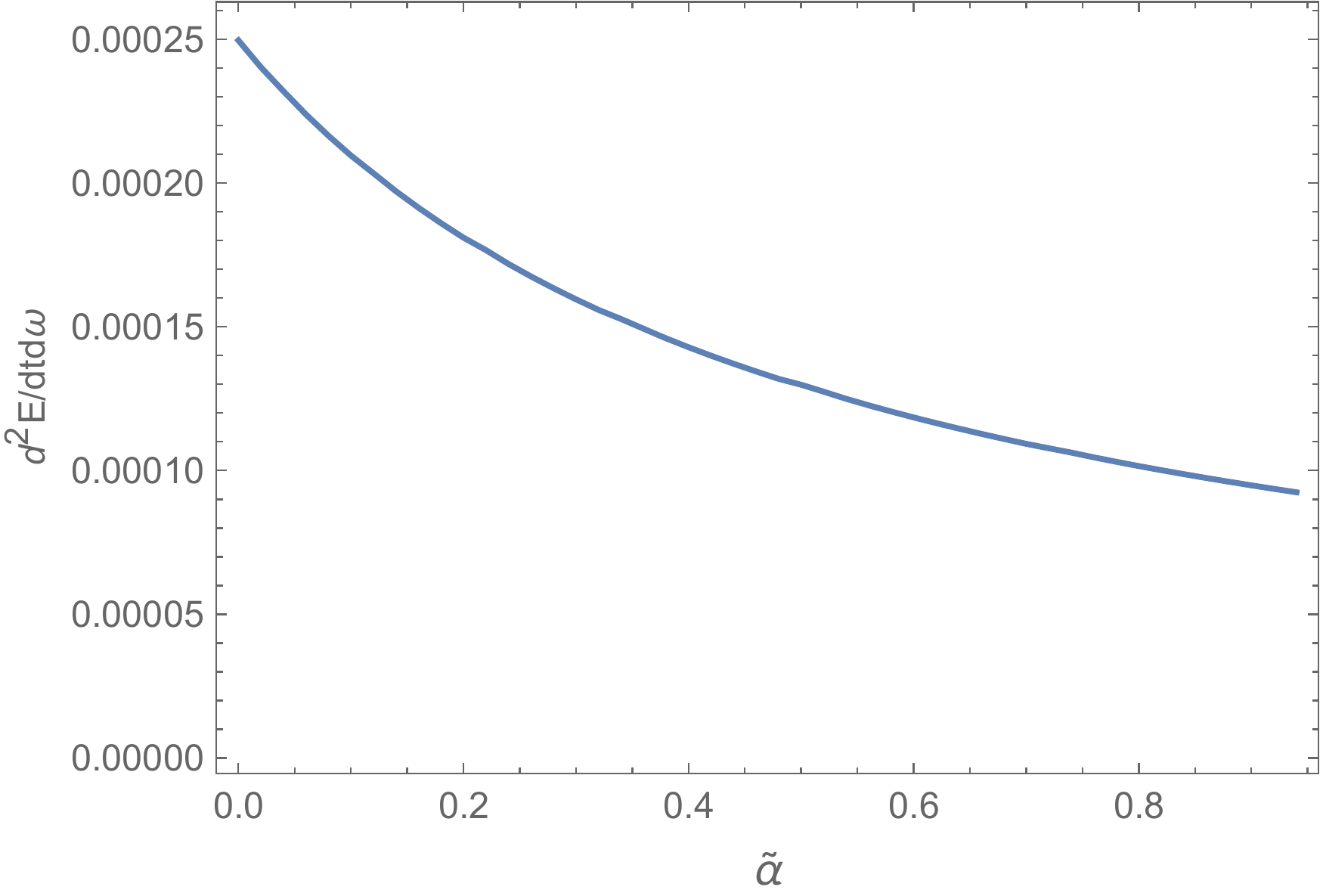}}\tabularnewline
\end{tabular}{\footnotesize\par}
\par\end{centering}
{\footnotesize{}\caption{\label{fig:PowerAlpha0} Effect of $\tilde{\alpha}$ on the power
spectra in the low energy limit. We fix $d=5,\Lambda=0.1,\xi=0$ here. The behavior in higher dimensional spacetime is qualitative similar.}
}{\footnotesize\par}
\end{figure}

Note that when $\xi=0$, the power spectra is nonzero when $\omega\to0$
due to the nonvanishing of greybody factor for dominant mode $l=0$.
It decreases monotonically with $\tilde{\alpha}$ as shown in Fig.
\ref{fig:PowerAlpha0}.

{\footnotesize{}}

\subsection{Effects of $d$ on the power spectra of Hawking radiation}

\begin{figure}[h]
\begin{centering}
{\footnotesize{}}%
\begin{tabular}{c}
{\footnotesize{}\includegraphics[scale=0.35]{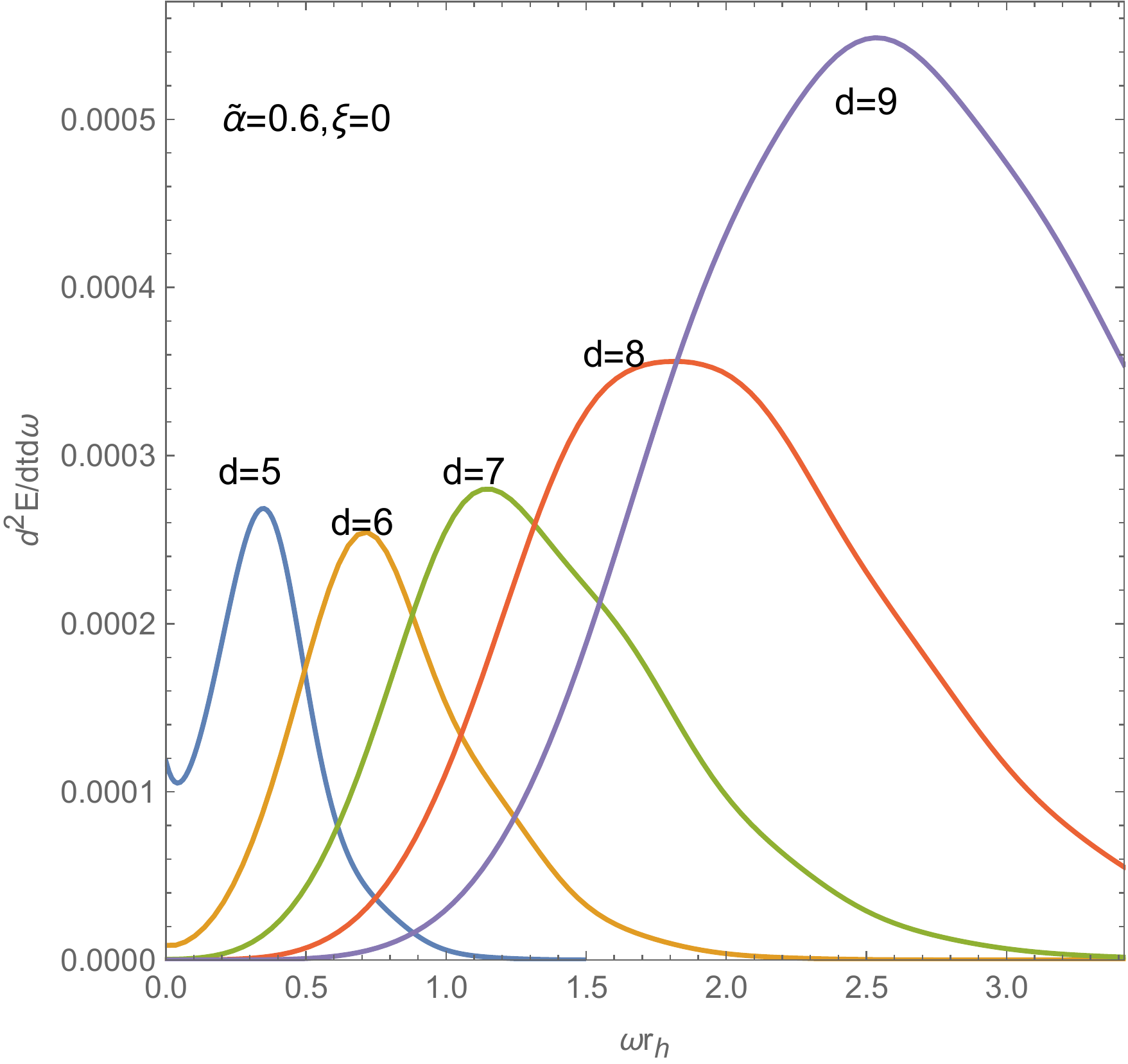}}\tabularnewline
\end{tabular}{\footnotesize\par}
\par\end{centering}
{\footnotesize{}\caption{\label{fig:PowerD} Effects of $d$ on the power spectra of Hawking
radiation. We fix $\Lambda=0.1$ here.}
}{\footnotesize\par}
\end{figure}

The effect of spacetime dimension $d$ on the power spectra of Hawking
radiation is shown in Fig. \ref{fig:PowerD}. Although the greybody
factor is suppressed by $d$ as shown in subsection \ref{subsec:dgreybody},
the significant increase of $T_{BH}$ leads to the overall enhancement
of the power spectra. This phenomenon has been found in \cite{Kanti2005,Harris2003,Pappas2016}.
We also see that the peak of the power spectra moves to higher frequency
in higher dimensions.

{\footnotesize{}}

\subsection{Competition between $\xi$ and $\Lambda$}

\begin{figure}[h]
\begin{centering}
{\footnotesize{}}%
\begin{tabular}{ccc}
{\footnotesize{}\includegraphics[scale=0.3]{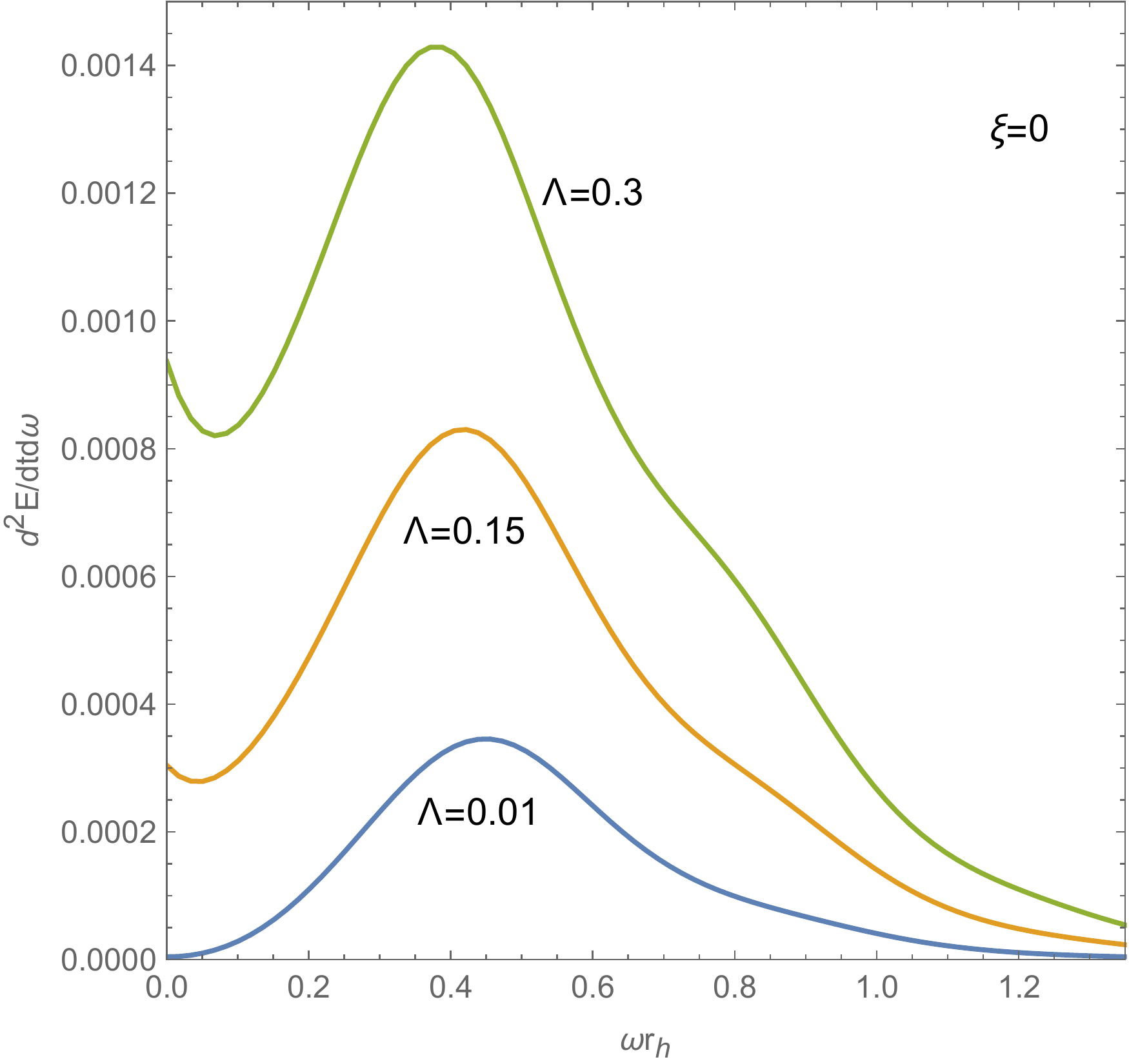}} & {\footnotesize{}\includegraphics[scale=0.3]{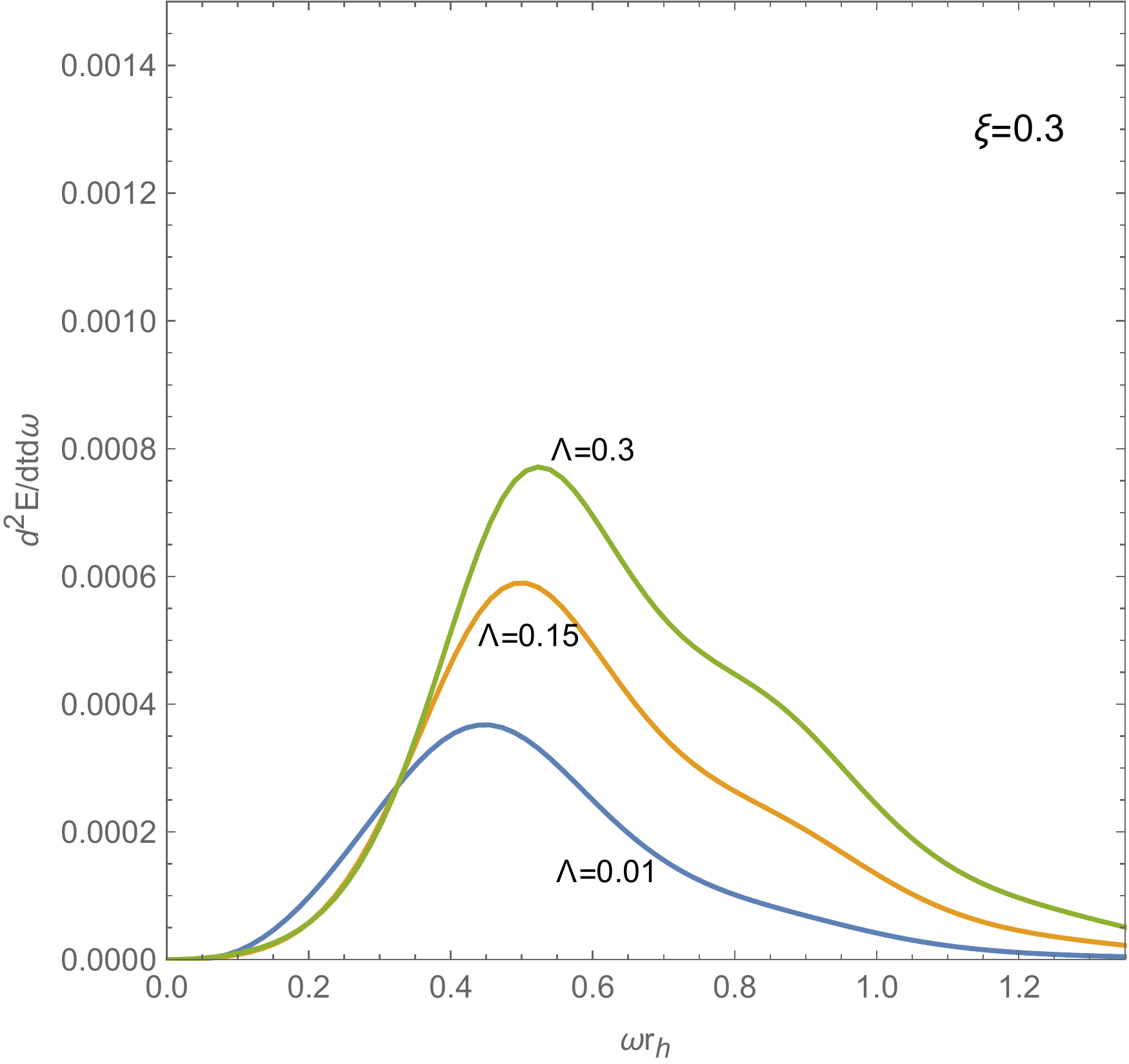}} & {\footnotesize{}\includegraphics[scale=0.3]{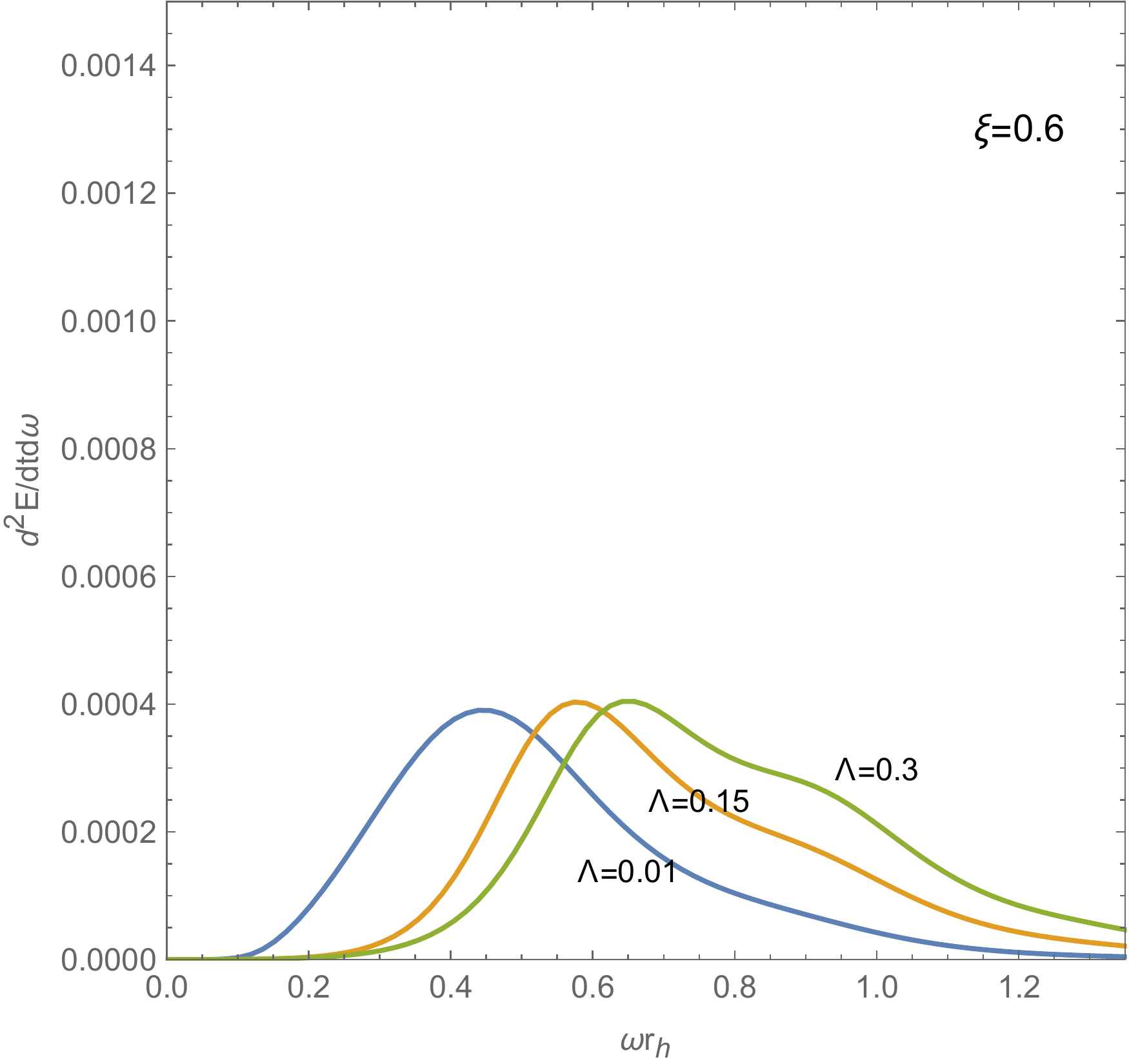}}\tabularnewline
\end{tabular}{\footnotesize\par}
\par\end{centering}
{\footnotesize{}\caption{\label{fig:PowerLambdaXi} Effects of $\Lambda$ on the {power spectra of Hawking radiation}.
We fix $d=5,\tilde{\alpha}=0.3$ here.}
}{\footnotesize\par}
\end{figure}

From Fig. \ref{fig:PowerLambdaXi}, we find that, when $\xi$ is small,
the power spectra is enhanced by $\Lambda$ in the whole energy regime.
For intermediate value of $\xi$, the enhancement occurs only in the
higher energy regime. When $\xi$ is large, besides the enhancement
in the high energy regime, the power spectra is actually suppressed
in the low energy regime as $\Lambda$ increases. This phenomenon has
been found for SdS black hole \cite{Pappas2016}. Note that the GB
coupling $\tilde{\alpha}$ does not change this behavior qualitatively.
The low energy behavior has also been discussed in \cite{Zhang2017}.

{\footnotesize{}}

\section{Summary \label{sec:Summary}}

We performed a comprehensive study of the greybody factors and power spectra of Hawking radiation
for the nonminimally coupled scalar field numerically in spherically
symmetric EGB-dS black hole spacetime. A recent work \cite{Zhang2017} had addressed the same problem in an approximate analytical
way, however, the validity of this method relied on the assumption that both the cosmological constant and the
nonminimal coupling were small. In this work, by numerically solving the scalar field equation and without making any approximation for the parameters we obtained the exact results for the greybody factors and
Hawking radiation spectra in the entire energy regime. The approximate analytical solutions \cite{Zhang2017} can be served as asymptotic
boundary conditions for our numerical integrations.

We first compared our results with the ones obtained by approximate analytical method
in \cite{Zhang2017}. In the low energy regime, these two set of results agree well for
small cosmological constant $\Lambda$ and nonminimal coupling constant
$\xi$. The deviations become large when $\Lambda$, $\xi$  or the energy of the emitted particle increase beyond the allowed regime.

The numerical results show that the greybody factor is suppressed
significantly by both the spacetime dimensions $d$ and the angular
momentum number $l$ of the scalar. The
GB coupling $\tilde{\alpha}$ always increases the greybody factor
 when other parameters are fixed. The effects
of scalar coupling $\xi$ on the greybody factor are relevant to the
GB coupling $\tilde{\alpha}$. However, when $\tilde{\alpha}$ is small, $\xi$ decreases the
greybody factor in the whole energy regime. When $\tilde{\alpha}$ is large,
$\xi$ decreases the greybody factor only in the low energy regime
but increases the greybody factor in the high energy regime instead.
This is different from the case of  Schwarzschild-de Sitter black hole where
$\xi$ always decreases the greybody factor \cite{Pappas2016}. The
effects of cosmological constant on the greybody factor are relevant
to scalar coupling constant: depending on the value of the nonminimal coupling constant, the positive $\Lambda$ can either help or hinder the particle
to overcome the potential barrier.
This phenomenon is due to the competition between the two roles $\Lambda$
plays: it subsidize energy to the particle and also the effective
mass of the particle.

We also analyzed the effects of $d,\xi,\tilde{\alpha},\Lambda$ on the energy
emission rate of Hawking radiation in detail. The energy emission
rate depends on both the greybody factor and the black hole temperature.
{For dS black holes, the definition of temperature is subtle. We compared
various temperature definitions and find that the normalized temperature is the most natural one for studying Hawking radiation. The normalized temperature decreases with $\alpha$. As a result, due to the dominant role of temperature on Hawking radiation,}
 we found that both $\tilde{\alpha}$
and $\xi$ suppress the power spectra,
unlike their subtle effects on the greybody factor.  Moreover,
the power spectra is overall enhanced significantly as $d$ increases
and the peak moves to higher energy regime. The effect of $\Lambda$
on the power spectra is relevant to the scalar coupling constant too.
When $\xi$ is small, it increases the power spectra in the whole energy
region. When $\xi$ is large, it increases the power spectra only
in higher energy regime and instead decreases the power spectra in
the lower energy regime.

\section{Acknowledgments}

P.-C. Li is supported by  NSFC Grant No. 11847241.  C.-Y. Zhang is supported by National Postdoctoral Program for Innovative
Talents BX201600005 and Project funded by China Postdoctoral Science Foundation.

\end{document}